\documentclass[superscriptaddress,twocolumn,showpacs,amssymb]{revtex4-1}
\usepackage{graphicx}
\usepackage[utf8]{inputenc}
\usepackage[english]{babel}
\usepackage{color}
\usepackage[T1]{fontenc}
\usepackage[utf8]{inputenc}
\usepackage{lmodern}
\usepackage{amsmath}
\usepackage{amssymb}
\usepackage{hyperref}
\usepackage[export]{adjustbox}
\usepackage{verbatim}
\usepackage{cases}

\newcommand{\dif}{\mathrm{d}}

 \newcommand{\im}{\mathbf{i}}
  
 \newcommand{\abs}[1]{\left\vert#1\right\vert}

\begin{document}

\title{Localization of soft modes at the depinning transition}

\author{Xiangyu Cao}
\email[]{xiangyu.cao08@gmail.com}
\affiliation{CNRS - LPTMS, Univ. Paris-Sud, Université Paris-Saclay,  France.}
\affiliation{Department of Physics, University of California, Berkeley, Berkeley CA 94720, USA.}
\author{Sebastian Bouzat}
\affiliation{CONICET - Centro Atomico Bariloche, 8400 S. C. de Bariloche, Argentina}
\author{Alejandro B. Kolton}
\affiliation{CONICET - Centro Atomico Bariloche, 8400 S. C. de Bariloche, Argentina}
\author{Alberto Rosso}
\email[]{alberto.rosso74@gmail.com}
\affiliation{CNRS - LPTMS, Univ. Paris-Sud, Université Paris-Saclay,  France.}

\date{\today}

\begin{abstract}
We characterize the soft modes of the dynamical matrix at the depinning transition, and compare it with the properties of the Anderson model (and long-range generalizations). The density of states at the edge of the spectrum displays a universal linear tail, different from the Lifshitz tails. The eigenvectors are instead very similar in the two matrix ensembles. We focus on the ground state (soft mode), which represents the epicenter of avalanche instabilities. We expect it to be localized in all finite dimensions, and make a clear connection between its localization length and the Larkin length of the depinning model. In the fully connected model, we show that the weak-strong pinning transition coincides with a peculiar localization transition of the ground state. 
\end{abstract}

\maketitle

\section{Introduction}
\begin{figure*}
\includegraphics[width=.8\textwidth]{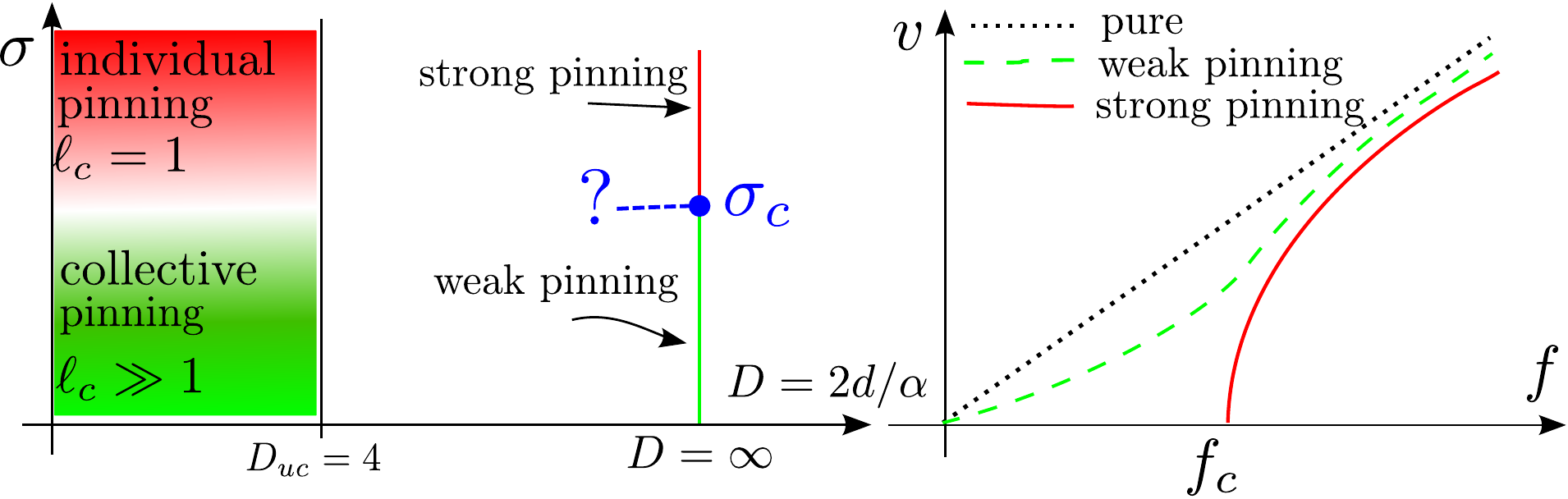}
\caption{\textit{Left}: Phase diagram for disordered elastic system within the Larkin approach. When $D <D_{uc} = 4$, pinning is always strong. By varying the disorder strength $\sigma$ results in a crossover between individual ($\ell_c \sim 1$) and collective ($\ell_c \gg1$) pinning regimes. When $D > D_{uc}$, $\ell_c$ diverges, the crossover disappears. In the fully connected model $D = \infty$, there is a transition between weak and strong pinning phases. We believe that such a transition does not exist for any finite $D$.  \textit{Right}: Velocity-force characteristics: for pure systems ($\sigma = 0$), $v \propto f$; for weak pinning ($\sigma < \sigma_c$), the velocity is non-zero for any $f > 0$; for strong pinning ($\sigma > \sigma_c$), there is a critical force $f_c > 0$, below which $v = 0$.}\label{fig:phase}
\end{figure*}

The presence of disorder is at the origin of  novel dynamical features that cannot be observed in pure systems. One of the most remarkable phenomena is the presence of \textit{avalanches}, namely, discontinuous and large re-organizations triggered by infinitesimal perturbations. Avalanches are observed in a host of experimental systems, ranging from the {Barkhausen} noise in ferromagnets~\cite{zapperi1998dynamics,DurinarXiv2016} to the propagation of a crack front~\cite{gao1989first,maaloy2006local,BSP08}, or the dynamics of the contact line in the liquid meniscus of a rough substrate~\cite{joanny1984model,rolley09}. All these cases are well understood in terms of the depinning of $d$-dimensional elastic interfaces in random media. Elasticity can be either short-range or involve long-range interactions, decaying as $\abs{r-r'}^{-(d+\alpha)}$, with $\alpha \in (0,2)$. It is shown that for a small drive, the interface is pinned in dynamically stable configurations. When the drive is above a finite threshold (called the critical force), the interface acquires a non-zero velocity. At the threshold, one observes scale-free avalanches~\cite{RLDW09}. 
Various arguments suggests that there is an upper-critical dimension, above which the depinning model has a mean field behavior. From a scaling analysis in terms of $D := 2d / \alpha$ (where $\alpha=2$ corresponds to short-range elasticity), one expects $D_{uc} = 4$. An anomalous behavior has been predicted at $D=\infty$ in a fully connected model~\cite{fisher1983threshold}. This model is known to display a transition, at a critical disorder strength $\sigma=\sigma_c$, between a strong pinning phase, characterized by a finite critical force, and a weak pinning phase, where there is no metastability and the interface cannot be pinned in the thermodynamic limit. The existence of such weak pinning phase at finite dimensions is still controversial (see Fig.\ref{fig:phase}). 

The configuration just before an avalanche is, by definition, marginally stable~\cite{Muller2015} and possesses soft-modes. In numerous situations, both in depinning ~\cite{Littlewood1986,Coppersmith1988,Middleton1992,Middleton1993} and models of amorphous and glassy materials~\cite{tsamados2009local,charbonneau16universal,Nandi2016} it is observed that the soft modes are localized, and the localization center is identical to the epicenter of {an} avalanche. In Ref.~\cite{Tanguy2004} the soft-mode localization in $d=1$ elastic interfaces was studied numerically as a function of $\alpha$. It was shown that, in a finite system and for $\alpha > 0$, the soft modes appear delocalized for weak disorder and localized for strong disorder. It remains to understand whether and for which range of $\alpha$ this is a genuine localization-delocalization transition or a finite-size crossover. In the first part of this paper, we address this question by studying the soft-modes of the marginally stable configurations, namely the ground-state and lowest excitations of their dynamical matrix. In particular we show the following:

\begin{itemize}
\item[-] When $D < D_{uc}$, the epicenter which triggers an avalanche instability is always localized.  At small disorder the pinning is collective, and the linear size of the epicenter can be identified with the \textit{Larkin length} of the elastic interface. 
\item[-] For the fully-connected model ($\alpha = 0, D = \infty$)~\cite{fisher1983threshold}the weak-strong pinning transition coincides with a localization transition of the ground state (soft mode) of the dynamical matrix. {\em The soft mode is localized in the strong pinning phase, and is delocalized in the weak pinning phase}. 
\end{itemize}

The previous results led us to relate the localization properties of the depinning dynamical matrix to the ones of the much simpler Anderson model in which the diagonal  disorder correlations are neglected.  This connection  is fruitful: we provide numerical evidences and analytical arguments  showing that the corresponding eigenvectors share the same localization features. However, the connection is not exact: The well-known 
Lifshitz tails~\cite{Lifshitz} of the eigenvalue distribution are not observed at depinning, where a simple linear tail is found (at any $D$). We can relate this behavior with the absence of the pseudo-gap at the depinning transition~\cite{Muller2015}.

In the last part of the paper we deal with the controversial question regarding the existence of a weak pinning phase at $ \infty > D > D_{uc}$. This phase should correspond to a delocalized soft mode in the dynamical matrix. In order to make progress in this difficult question we study instead the ground-state properties of the long-range hopping generalization of the Anderson model~\cite{rodriguez03anderson}.
Using a novel and extensive numerical study of this model, together with a generalization of the Lifshitz argument,  we provide strong evidence that the ground state is always localized by rare potential valleys, but has peculiar properties reminiscent of the fully connected depinning soft mode.

The paper is organized as follows. In section \ref{sec:model}, we define the disorder elastic models, review the Larkin approach and introduce the dynamical matrix. In section \ref{sec:SR}, we study numerically the $1d$, short-range case, and show that the Larkin length coincides with the localization length of the ground state of the dynamical matrix.  In section \ref{sec:mf}, we study the ground state of the fully connected depinning model.In section \ref{sec:LR}, we focus on the ground state of the long-range hopping Anderson model using 
a novel numerical technique and the Lifshitz argument. We discuss open questions and perspectives in section \ref{sec:conclusion}. The main text is complemented by a few appendices. 

\section{Depinning transition: generalities}\label{sec:model}

\subsection{The model}
\begin{figure}
\includegraphics[width=.7\columnwidth]{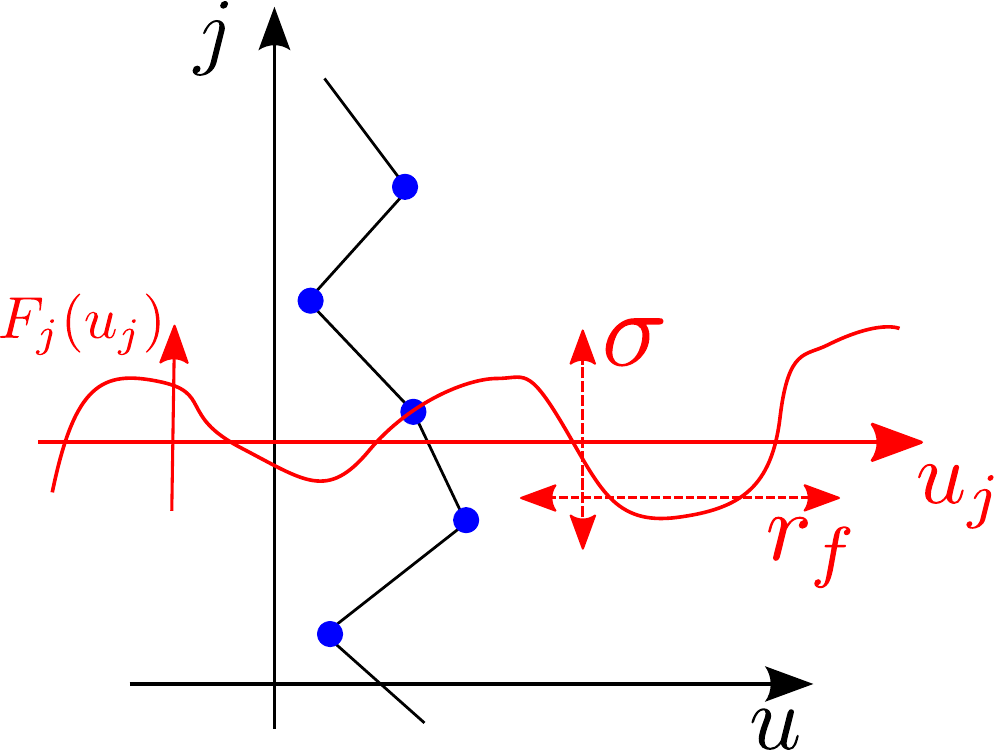}
\caption{An illustration of the disordered elastic model in the $d=1$ case. Each point $j$ has its own disorder force $F_j(u_j) = -\partial_{u_j} V_j(u_j)$ as a function of $u_j$. We depict one of them, and indicate the parameters characterizing its correlation length $r_f$ and its magnitude $\sigma$, see Eq. \eqref{eq:correlator}.}\label{fig:interface}
\end{figure}

We model a $d$-dimensional interface embedded on a $d+1$ disordered material as a collection of blocks, located at each site of a $d$-dimensional regular lattice $i=1,\ldots,L^d$, and characterized by a continuous displacement $u_1,\ldots,u_{L^d}$ in the $d+1$ transverse direction. An illustration for the $d = 1$ case is provided in Fig. \ref{fig:interface}.

The interface energy reads:
\begin{equation}
E = \frac{1}{2} \left( \sum_{i,j} G_{ij} \, (u_i-u_j)^2\right) + \sum_i V_i(u_i) \,.
\end{equation}
Here $ V_i(u_i)$ is the pinning potential, characterized below. The quadratic term accounts for the elasticity, with 
$G_{ij}$ being the spring constant associated with the blocks $i$ and $j$. The Hookian matrix $G$ is then symmetric and its non-diagonal elements are all positive. We set $G_{ii}=-\sum_{j\ne i} G_{ij}$ so that each row sums to zero. In this work $G$ takes the form of a  fractional Laplacian $-(-\nabla^2)^{\alpha/2}$, with $\alpha \in [0,2]$. When $\alpha=2$, it coincides with the standard Laplacian, when $\alpha=0$, it represents a fully connected limit with spring constants $G_{ij} =1/L^d$. For $\alpha \in (0,2)$ the fractional Laplacian is long ranged and the spring's strength decays with the distance between blocks in a $1/r^{d+\alpha}$ fashion. The fractional Laplacian has a simple definition in terms of its Fourier transform ~\cite{zoia}. With periodic boundary conditions and on a discrete lattice it reads:
\begin{equation}
\label{eq:frac}
G^{(\alpha)}_{ij} = -\frac{1}{L}\sum_{k=1}^{L-1}  e^{\im 2\pi k(i-j)/L} (2 - 2\cos(2\pi k / L))^{\alpha/2} \,,
\end{equation}
where the term $-(2(1-\cos(q)))^{\alpha/2}$ is the simplest regularization of the fractional Laplacian, $-|q|^\alpha$, on the lattice. We will often omit the superscript $(\alpha)$ and write simply $G_{ij} = G^{(\alpha)}_{ij}$. Properties of the fractional Laplacian are reviewed in Appendix~\ref{app:frac}.

The interface is pulled by an external force $f$ and, in the over-damped limit, its Equation of motion is given by: 
\begin{equation}\label{eq:eqmotion}
\dot{u}_j(t)=-\frac{\delta E}{\delta u_j(t)} +f = \sum_{k=1}^{L^d} G_{jk} \, (u_k- u_j) + F_j(u_j) + f
\end{equation}
Note that the pinning force, $F_j(u) = -V^{'}_j(u)$, represents the only non-linear term of Eq.(\ref{eq:eqmotion}) and in its absence the interface is flat and moves with velocity $\dif u_j / \dif t=f$.
On the contrary, the numerical solution of Eq. (\ref{eq:eqmotion}) displays a richer behavior: for small forces the interface  is pinned in metastable configurations and can slide only above some sample-dependent critical force $f_c(L)$. The existence of a critical force that remains finite in the thermodynamic limit, $f_c:=f_c(L\to \infty)$, was first argued by Larkin for short range elastic systems in $d<4$. Here we recall his discussion, within the more general context of the fractional Laplacian.

\subsection{The Larkin approach} \label{sec:Larkin} 
To make progress we first write the correlations of the pinning force as
\begin{equation}
\label{eq:correlator}
\overline{F_{i}(u) F_j(u')}= \sigma^2 \Delta(u-u')\delta_{ij} \,. 
\end{equation}
with  $\Delta(0)=1$, $\sigma$  the strength of the pinning force and the symbol $\overline{ \cdots} $ stands for the average over disorder realizations. Two cases are physically relevant: (i) either the function $\Delta$ is short-ranged, as for random bond or random field disorder, (ii) either the function $\Delta$  is periodic, as for charge density waves. In both cases $r_f$ sets the scale of the distances along $u$ between consecutive zeros of the pinning force $F(u)$, see Figure \ref{fig:interface}.
Larkin introduced a dramatic simplification taking $r_f \to \infty$, so that the random force $F_j$ does not depends on $u$. This toy model can be solved analytically: in the long time limit the interface slides with a finite velocity, $v=f +  \langle F \rangle $ (with $\langle F \rangle =(\sum F_j)/L^d$), and a time-independent shape, $\tilde u_i$,  which obeys to the following Equation, 
\begin{equation}
\label{eq:Larkinmodel}
\sum_{k=1}^{L^d} G_{jk} (\tilde u_k- \tilde u_j) =  \langle F \rangle - F_j \, .
\end{equation}
Two different regimes should be distinguished as a function of the \textit{effective dimension} $D$, which is determined in turn by the spatial dimension $d$ and the long-range exponent $\alpha$ via~\footnote{Note that the definition of $D$ is chosen out of convenience. Any definition of $D$ would be legitimate provided: $D = d$ in the short range $\alpha = 2$ case, and $D = \infty$ in the $\alpha = 0$ mean field case, and $D_{uc} = 4$ is upper critical. Moreover not all universal quantities of the depinning model can be expressed in terms of $D$ solely, for example the Larkin exponent Eq. \eqref{eq:Larkinlenght}.}
\begin{equation}
D = 2 d / \alpha \,. \label{eq:EffectiveD}
\end{equation}
When $D < 4$, the roughness of interface (denoted as $B(r)$ below), i.e., its displacement fluctuation in the $u$-direction as a function of separation in the $r$-direction, is characterized by an exponent $ \zeta_L$:
\begin{equation}
\label{eq:Larkinlenght}
B(r)=\overline{(u_r-u_0)^2} = \sigma^2 r^{2 \zeta_L} \;\,  \text{with} \,\,  \zeta_L= \alpha-d/2 >0 \, .
\end{equation}
When $D > 4$, the wandering of the interface remains bounded and $\zeta_L=0$. 

It is tempting to interpret these results with the phase diagram in Figure \ref{fig:phase} (Left panel) and to identify the upper-critical dimension with 
\begin{equation} D_{uc} = 4 \,. \label{eq:Duc} \end{equation}
In particular, for the short-range case, $\alpha=2$, this corresponds to the upper critical dimension $d = 4$. For the 1D interface, $d = 1$, Eq. \eqref{eq:Duc} gives a lower critical long-range exponent $\alpha_c = 1/2$.
\begin{itemize}
\item When $D <  D_{uc}$, the model is in the \textit{strong pinning} phase. Disorder is  relevant and
the critical force has a  finite value when $L\to \infty$.  A natural scale, $\ell_c$, called Larkin length,  is associated with the Larkin breakdown, namely  $B(\ell_c)=r_f^2$, so that
\begin{equation} 
\label{eq:Larkin}
 \ell_c= \left( \frac{r_f}{\sigma}\right)^{1/\zeta_L} \, .
\end{equation}
Note that at large scales $\gg \ell_c$ the Larkin exponent do not describe the roughness of the interface.
 However, in the {\em collective pinning regime} (see Figure \ref{fig:phase}, Left panel), namely when $\ell_c$ is much larger than the lattice spacing or the distance between impurities, one can use the Larkin model to provide a good estimation of the critical force \cite{L70,larkin1979pinning,agoritsas2012disordered,patinet2013quantitative,Demery2014b,demery2014microstructural,fyodorov2017exponential}:
\begin{equation} 
\label{eq:Larkin_force}
f_c \sim \frac{\sum_{j=1}^{\ell_c^d} F_j}{\ell_c^d} \approx \frac{\sigma}{\ell_c^{d/2}} \sim \left(  \frac{\sigma^{2 \alpha}}{r_f^d}\right)^{\frac{1}{2 \alpha-d}} \,.
\end{equation}
When the driving force $f < f_c$, the interface is pinned, $v = 0$, see Figure \ref{fig:phase}, Right panel.
 \item When $D > D_{uc}$, the interface becomes flat and the mean field description becomes correct. There is less consensus on the phase diagram in this region. D. Fisher provided an analytical solution for the fully connected model with $\alpha=0$~\cite{fisher1983threshold,fisher85}. As we will review in section \ref{sec:mf}, this model has two phases, separated by a critical strength of disorder $\sigma_c$. When $\sigma < \sigma_c$, the system is in the \textit{weak pinning} phase. Disorder is irrelevant and the critical force vanishes in the thermodynamic limit, $f_c = \lim_{L \to \infty} f_{c,L} = 0$. At small force, the interface slides with a velocity proportional to the force, but with a very small proportionality constant (see Figure \ref{fig:phase}, Right panel). As $\sigma > \sigma_c$, the system is in the {\em strong pinning} phase and display a genuine depinning transition: $f_c > 0$. It is still an open question to establish if the weak pinning phase is a peculiarity of the fully connected model, or holds for all $D>D_{uc}$. The results presented below are in favor of the former possibility.
\end{itemize}

\subsection{Depinning soft modes: the dynamical matrix}\label{sec:matrix}
Let us consider the protocol in which the elastic interface is driven quasi-statically by increasing the force $f$ towards the depinning critical point $f_c$.  Upon an infinitesimal increase of the drive $f \to  f + \delta $, there can be two possibilities: (i), the new metastable configuration differs from the previous one only by a small amount $\delta u_j \propto \delta $; (ii), the new metastable configuration differs from the previous one by a finite amount: this is called an \textit{avalanche}~\cite{RLDW09}. In the latter case, the increase $\delta$ needed to trigger an avalanche is known to be 
\begin{equation} \delta \propto 1/L^d \label{eq:deltafmin} \end{equation} for all depinning models; this fact is also known as the absence of pseudo gap~\cite{Muller2015}. The configurations just before the avalanche is called {\em marginally stable}, and the difference between the two consecutive configurations defines the avalanche size $S$. The size distribution is found to be $P(S) \propto S^{-\tau}F(S / S_{\text{max}})$, \textit{i.e.}, it is a power-law with a cut off $S_{\text{max}}(f)$, which diverges as $|f-f_c|^\chi$ near the depinning transition. Both the exponents $\tau$ and $\chi$ are universal (they depend only on $d$ and $\alpha$), and can be calculated by functional renormalization group in a second order $\epsilon = D_{uc}-d$ expansion \cite{CGLD98,Chauve2001,Chauve-PRB2000}. Above $D_{uc}$ they saturate at the mean field value $\tau=3/2$ and $\chi=2$. 

Around each metastable configuration $u^{(0)}$, we may linearize Eq.(\ref{eq:eqmotion}). By writing $u_j(t)=u^{(0)}_j+\delta u_j(t)$  and using the metastability condition  $\sum_k G_{jk} u^{(0)}_{k} + F_j(u^{(0)}_j) +f)=0$ we get:
\begin{eqnarray}\label{eq:eqmotiondiscrete}
\delta {\dot u}_j(t) =  -\sum_k M_{jk} \delta {u}_k
\end{eqnarray}
where  the \textit{dynamical matrix} $M$ (also known as the \textit{Hessian}) is defined as
\begin{equation}
M_{jk} = \begin{cases}
 - F'_j(u^{(0)}_j) - G_{jj} & j = k \\
 -G_{jk} & j \neq k \,.
\end{cases}
\label{eq:M}
\end{equation}
The matrix $M$ is real and symmetric, so its eigenvalues are all real, and we denote them in the ascending order by $\lambda_0 \leq \lambda_1 \leq \dots$. The meta-stability of the configuration implies that $\lambda_0 \geq 0$. In the presence of marginal stability, the lowest eigenvalue vanishes $\lambda_0 = 0$. 
However, $\lambda_0 = 0$ is not a sufficient condition for an avalanche: indeed, in the pure system ($F_j = 0$), $\lambda_0 =0$ always holds, and this corresponds to the translation invariance. Therefore, it is important to study the \textit{ground state} $(\phi_j)_{j=1}^{L^d}$, defined as the eigenvector of $\lambda_0 = 0$:
\begin{equation}
\sum_{k} M_{jk} \phi_k = 0 \,.
\end{equation}
The ground state, which we also call \textit{the soft mode}, identifies the marginally stable direction. In particular, the properties of the matrix $M$ ensure that the coefficients $\phi_j$ are all positive (see Appendix \ref{sec:props}). This fact can be also understood in light of the Middleton theorem \cite{MiddletonPRL,Middleton1993}, that ensures that if $G_{jk} \geq 0$, then independently of the initial condition, after a transient, the interface moves only in the forward direction. 
 
We propose that two scenarios are possible: (i) if $\phi_j$ is evenly distributed in the entire system ($\phi_j \sim 1 / L^{d}$), there are no avalanches; if this remains true in the $L \to \infty$ limit, we are in the weak pinning phase; (ii) if $\phi_j$ is localized in a finite portion of the system, it then represents the epicenter of an avalanche. 

\section{Short range depinning}\label{sec:SR}
In this section, we focus on the short range ($\alpha = 2$) case and generalize our results to $D<D_{uc}$.

\subsection{Numerical set up} 
In order to  sample metastable configurations close to the depinning transition, it is convenient to study a variant of the elastic model Eq. \eqref{eq:eqmotion}:
\begin{equation}\label{eq:eqmotion_shortrange}
\dot{u}_j(t)= (u_{j+1} + u_{j-1} - 2 u_j) + F_j(u_j) + m^2(w-u_j) \,.
\end{equation} 
 Here we replace the constant force $f$ with a soft spring force $m^2(w-u_j)$. When $m \gtrsim 1/L$, the metastable states of Eq. \eqref{eq:eqmotion_shortrange} display the same statistical properties of the metastable states at a force $f$ slightly below the critical force: $f_c - f \propto m^{-1/\nu}$ ($\nu$ is the critical exponent of the length scale diverging at the depinning transition)~\cite{RLDW09,kolton2013}. To find the metastable configurations, we select an increasing sequence of $w$, for each of which we target the first metastable state using an efficient algorithm~\cite{werner02rough}. 
 The random potential $V_j(u)$ is obtained by interpolating, with cubic spline, a sequence of uncorrelated Gaussian random numbers with zero mean and variance $\sigma^2 r_f^2$, assigned to evenly spaced point $u = 1, 2, \dots$, so that $r_f = 1$~\cite{werner02rough,Ferrero2013CR}. Therefore, $V_j(u)$ has continuous second derivative, and $F'_j(u_j^{(0)})$ in Eq.~(\ref{eq:M}) is well-defined. 
 Every statistics below is performed by averaging over more than $1000$ uncorrelated configurations that are visited in a long enough run, with $m = 1/L$.

\begin{figure}[h]
\includegraphics[width=.9\columnwidth]{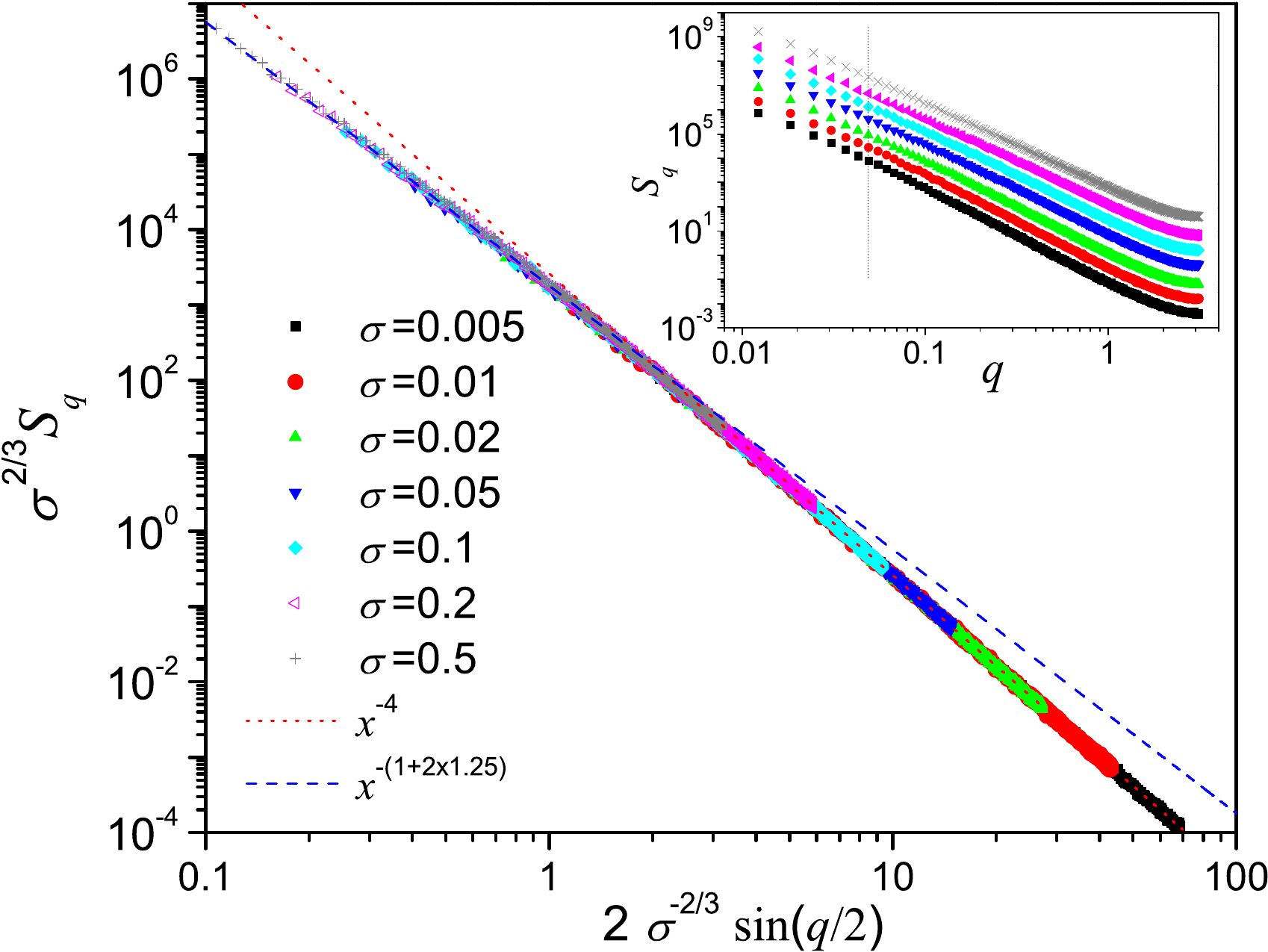}          
\caption{Rescaled averaged structure factor of the critical configuration, as a function of disorder, for $\sigma<1$. Dashed lines indicate two regime of roughness: at small length-scales we find the Larkin regime 
$S_q\sim q^{-4}$, and at large length-scales $S_q\sim q^{-(1+2\zeta)}$ with $\zeta\approx 1.25$ the random-manifold exponent at depinning. The crossover length corresponds to the Larkin length 
${\ell}_c \sim \sigma^{-2/3}$. Inset: raw data.
\label{fig:sq}}
\end{figure}   
As a test of the numerical method, we illustrate the significance of the Larkin length ${\ell}_c$, defined in Eq. \ref{eq:Larkin}, in terms of the shape of the metastable configurations. We look at the structure factor, defined as:
\begin{equation}
S_q \equiv \overline{|u_q|^2} \,,\, u_q :=  \sum_j u_j e^{\im q j} \,,
\end{equation}
with $q= 2 \pi n /L$,  for $n=-L/2, ...,L/2-1 \,.$ Following Larkin's ideas, one expects  that the interface is described by the Larkin model [eq. \eqref{eq:Larkinlenght}] up to scale $\ell_c$, beyond which it displays the large-distance depinning roughness. Thus, the structure factor is expected to satisfy the scaling behavior:
\begin{equation}
S_q \sigma^{\frac23} = \widetilde{H}\left(\frac{q}{q_c}\right) \,,\,   \widetilde{H}(x) \sim \begin{cases}
x^{-(1+2\zeta)} & x \ll 1 \,, \\ 
x^{-4} & x \gg 1 \,,
\end{cases}  \label{eq:Sq}
\end{equation} 
where $q_c = 2 \pi / \ell_c \sim \sigma^{2/3}$, and $\zeta \approx 1.25$ is known numerically~\cite{Ferrero2013}. The collapse of  $S_q$ for different disorder strengths $\sigma < 1$ is shown in Fig.~\ref{fig:sq}, where we observe a single master curve agreeing with Eq. \eqref{eq:Sq}. 

\begin{figure}[h]
\includegraphics[width=1\columnwidth]{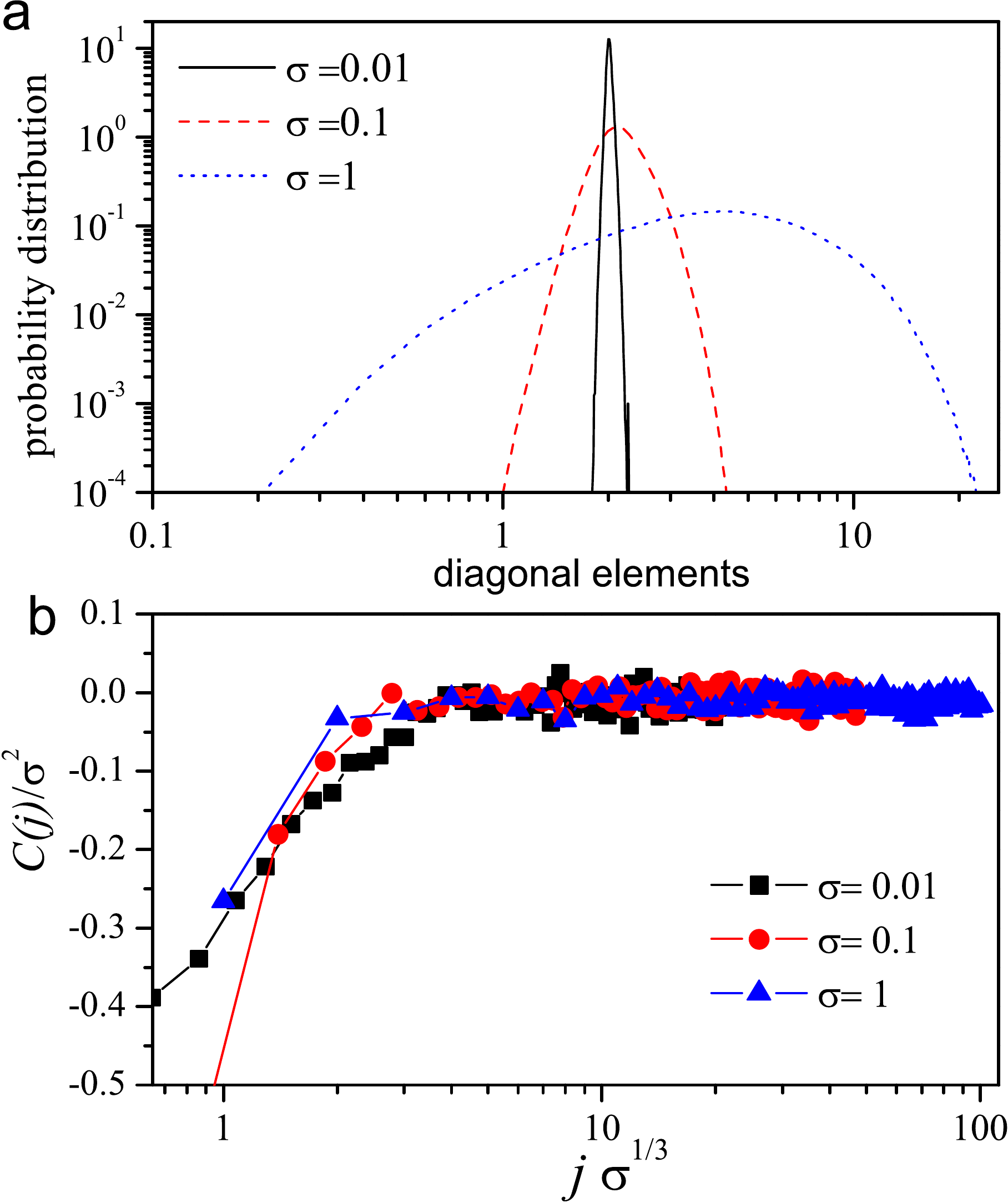}
\caption{(a) Probability distribution of the stability matrix diagonal elements as a function 
of the disorder strength $\sigma$. (b) The covariance  of  diagonal elements $C(|i-j|) := \overline{M_{ii}M_{jj}}^c.$
\label{fig:diagonal}}
\end{figure}

From now on we study the dynamical matrix, defined by Eq. \eqref{eq:M}.  The distribution of its diagonal elements displays a non-trivial shape, as shown in Fig. \ref{fig:diagonal}(a). It becomes peaked around $-G_{jj} = 2$ for small disorder $\sigma$. Their covariance turns out to be negative, up to a correlation length scaling roughly as $\sigma^{-1/3}$. 

\begin{figure}[h]
(a)\includegraphics[width=1\columnwidth,valign=t]{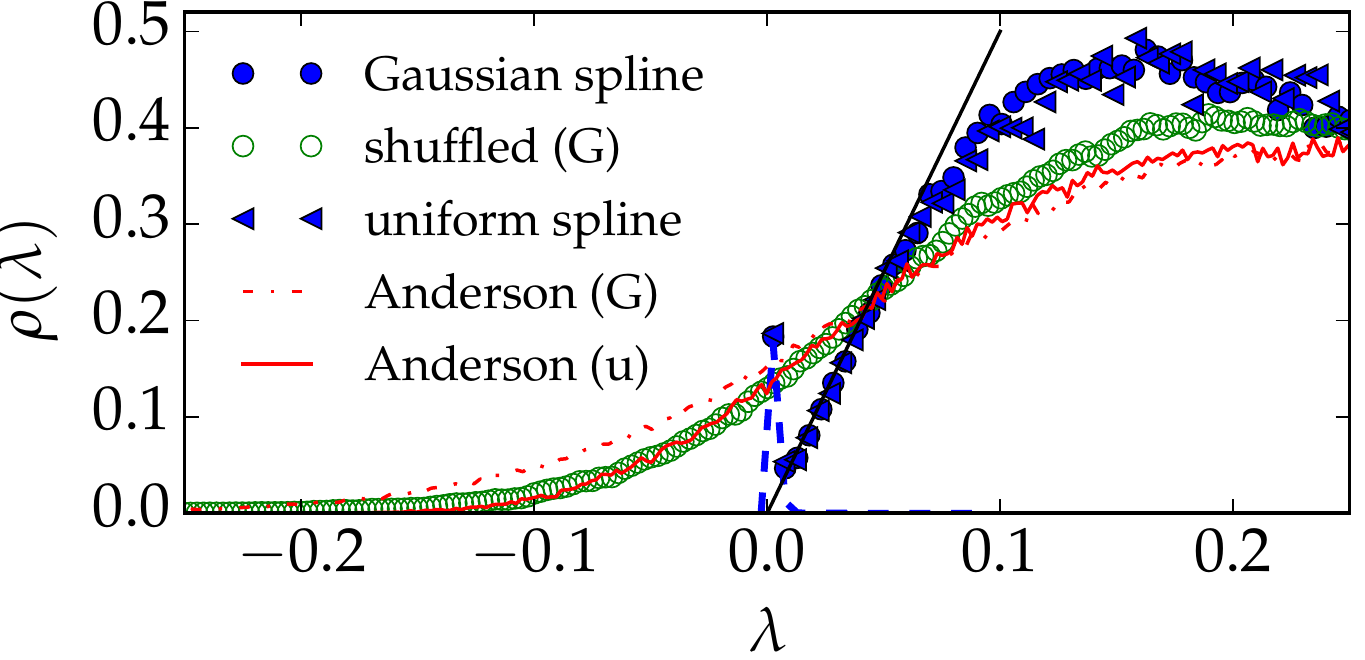}
(b)\includegraphics[width=1\columnwidth,valign=t]{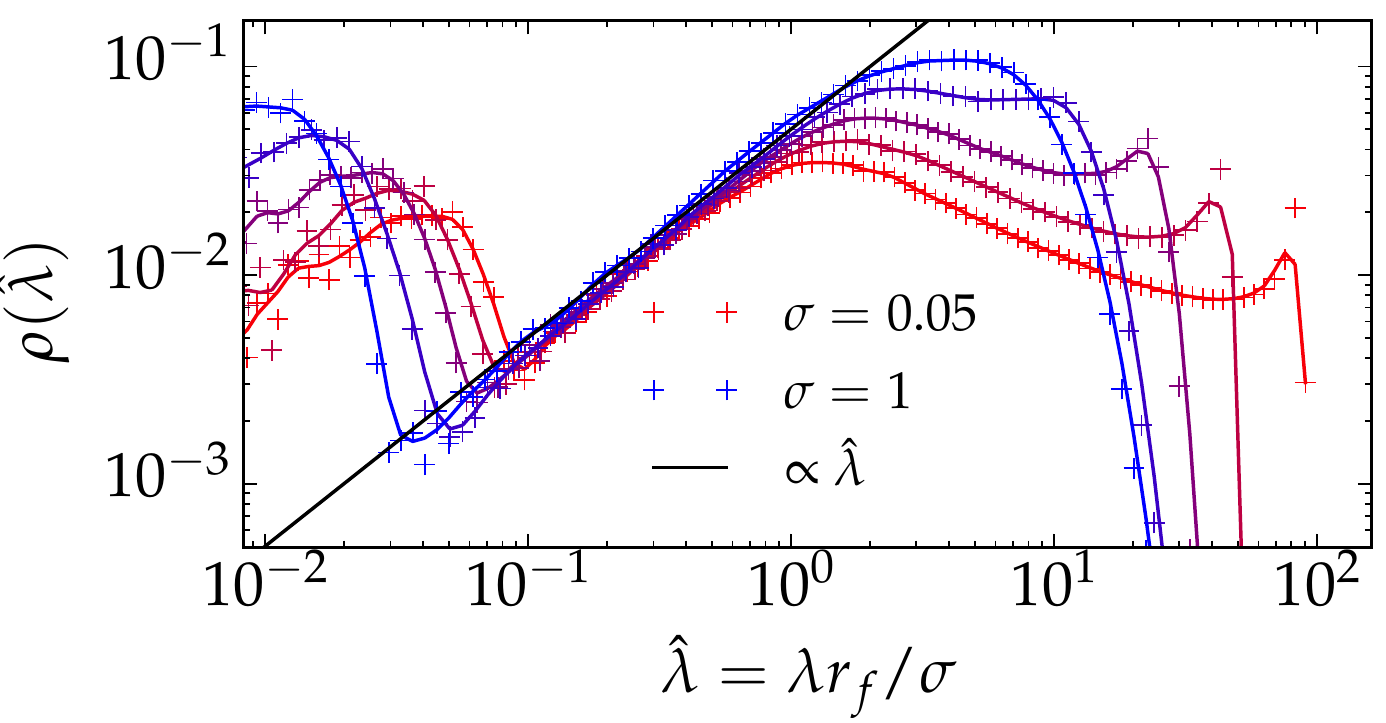}
\caption{(a) Filled markers: Density of states (DoS) of dynamical matrix of marginally metastable configurations (before an avalanche of size $S > 1$), in a disordered potential whose values at $u_j = 1,2,\dots$ are normally (circles) or uniformly (triangles) distributed, with variance $\sigma^2 = 0.1^2$ for both. The thick dashed curves around $\lambda = 0$ show the contribution of the ground state $\lambda_0$ to the DoS. The black line is a linear fit $\rho(\lambda) = 5 \lambda$. Empty markers: DoS of dynamical matrices with  diagonal elements randomly shuffled. Solid (dashed) red curve: 1d Anderson model with uniform (Gaussian) diagonal elements, with the same mean ($0.345$) and standard deviation ($0.225$) of the dynamical matrix diagonal elements. (b) Distribution of rescaled eigenvalues $\hat{\lambda} = \lambda r_f / \sigma$ of the dynamical matrix, with different $\sigma=1, 0.5, 0.2, 0.1, 0.05$. The solid curves are guides to the eyes. The straight line is a linear fit $\rho(\hat{\lambda}) = \hat{\lambda}/20.$ The histograms are binned in log scale, resolving the peak at $\lambda = 0$.
\label{fig:cumulativedistribution}}
\end{figure}   

\subsection{Analogy with the Anderson model }\label{sec:anderson}
The properties of the ground state of the dynamical matrix are not easy to predict because the statistics of the diagonal elements $M_{jj} = 2 d +  W_j$, where  $W_j = -F'_j(u^{(0)}_j)$, is generated by the complex dynamics of the interface and displays non-trivial correlations [recall Fig.~\ref{fig:diagonal} (b)]. In particular, its ground state energy is zero. However it is always instructive to discuss the case where the $W_j$ are independent and identically distributed uniform random variables. Within this approximation, the dynamical matrix takes the form of the Anderson model:
\begin{equation} M_{ij} = \begin{cases}
2d +W_j & i = j \; \text{with} \; W_j \in [0, W_d] \\
-1 & \text{$i, j$ nearest neighbourgs } \\
0 & \text{otherwise} \,,\, \label{eq:MijAnderson}
\end{cases} \end{equation}
where $W_d$ controls the disorder strength. Many results are known for this model. For $d \le 2$, all eigenvectors are localized: they are concentrated around a localization center $j_{\max}$ and decay with a characteristic \textit{localization length}, $\xi_{\text{loc}}$. For $d>2$ a sharp transition between localized and delocalized eigenvectors occurs in the bulk of the spectrum.

At the lower edge, and \textit{a fortiori} for the ground state, a different behavior appears and is sensible to the extreme fluctuations of the disorder. In particular, it was known since Lifshitz~\cite{Lifshitz} that the density of states near the edge develops a non-perturbative tail. More recently, it is proved that in any dimension \cite{Klopp2002}, all the states at the edge are  localized, and display an exponential far tail. This feature can be understood by modeling the lowest valley of the disordered potential with a single attractive impurity: $V(x) = -v_0 \delta (x)$ ($v_0 > 0$). The bound ground state of the corresponding Schroedinger Equation is well-known to be exponential at large distance; in particular, in 1d, we have $\phi(x) \propto e^{-v_0 \abs{x} / 2 }$. 

It is possible to generalize the Anderson model to the case of long-range hopping~\cite{rodriguez03anderson}:
\begin{equation} 
M_{ij} = \begin{cases}
W_j - G_{jj}^{(\alpha)}  & i = j  \\
-G_{ij}^{(\alpha)} & \text{$i\neq j$} \,, \\
\end{cases}  \label{eq:rodriguez}
\end{equation}
with $G_{ij}^{(\alpha)}$ being the fractional Laplacian, defined in Eq. \eqref{eq:frac}. This model has the same form of the depinning dynamical matrix in presence of long range elastic interactions, but similarly to the Anderson model,  the $W_j$ are independent and identically distributed uniform random variables.

In the following, we study numerically the eigenvalues and eigenvectors of the dynamical matrix at the depinning transition. To compare these results with the Anderson model, we fix the disorder strength of the latter as
\begin{equation}
W_d = \sqrt{12} \times \mathrm{std}\left[F'_j(u^{(0)}_j)\right] \sim \sigma/r_f \label{eq:Wdmatch}
\end{equation}
so that the diagonal elements of the two matrix ensembles have the same variance.

\subsection{Depinning density of states}
We now turn to the left-tail of the density of states (DoS) $\rho(\lambda), \lambda \sim 0$. To motivate the study of this quantity, we recall its close relation to the low-frequency vibration spectrum of the \textit{undamped} dynamics (that is, replacing $\dif / \dif t$ by $\dif^2 / \dif t^2$ in the Equation of motion \eqref{eq:eqmotiondiscrete}) near a stable Equilibrium. The frequencies $\omega_j$ of the approximate harmonic system are related to the eigenvalues of the dynamic matrix by 
\begin{equation} \omega_j = \sqrt{\lambda_j} \,. \label{eq:omega} \end{equation}
The spectrum of $\omega_j$ is experimentally accessible in crystalline and amorphous solids. Its peculiar features in the latter, such as the boson peak~\cite{philips2012amorphous}, are a subject of active investigation. In this context, it is interesting to study the question in the depinning case, which is quite unique in the realm of disordered systems.

For this, we first diagonalize the 1D, short-range dynamic matrices obtained above with standard numerical routines and compute the DoS by averaging over several realizations. The result, shown in Fig.~\ref{fig:cumulativedistribution}, is a \textit{linear} spectrum of lowest excitations:
\begin{equation}  \rho(\lambda) \sim \mathrm{c}\, \lambda \,,\,  0 <  \lambda \ll 1 \,,\, \mathrm{c} \sim \sigma^{-2} r_f^2   \,. \label{eq:linearDoS} \end{equation} 
For configurations just before an avalanche instability, we also observe a sharp peak at $\lambda = 0$ of a vanishing amplitude $\sim 1/L$, contributed by the marginally stable ground state $\lambda_0$.

Let us compare the depinning result Eq. \eqref{eq:linearDoS} with what is known in the uncorrelated Anderson model. In the absence of disorder, $\rho(\lambda) \propto \lambda^{d/2-1}$ for $0 < \lambda \ll 1$; the disorder shifts the left limit of the DoS $\lambda = 0 \leadsto \lambda_0$, and changes qualitatively the algebraic behavior at the edge. In general, the modified behavior depends on the disorder distribution; in particular, for uniform distribution $W_j \in [a,b]$, we have the Lifshitz tail~\cite{Lifshitz,bookonlocalization}
\begin{equation}  \rho(\lambda) \sim (\lambda - \lambda_0)^{d/2 - 1} \exp\left[- C (\lambda  - \lambda_0)^{-d/2}\right] \label{eq:lifshifts} \end{equation} 
where $C$ is a constant depending on the disorder and $\lambda_0 = a$. However, the linear tail of dynamical matrix is \textit{not} originated by the diagonal elements' distribution in Fig.~\ref{fig:diagonal}. Indeed, if we shuffle them randomly, the linear tail is destroyed and replaced by a Lifshitz tail, as we show in Fig.~\ref{fig:cumulativedistribution} (a). 

We believe that Eq. \eqref{eq:linearDoS} is a universal fingerprint of the depinning transition in all dimensions, as it is the spectrum of the soft spot excitations. To justify this claim, we recall from Eq. \eqref{eq:deltafmin} that in order to trigger an avalanche, the extra force one needs to apply is $ \delta  = m^2 \delta w \propto 1/L^d$. Since the disorder potential is smooth, we write the effective potential acting on the soft spot near instability as $\tilde{V}(u) = -u^3/3 + \delta \times u$~\cite{middleton92rounding}. Its stable position is $u_* = -\sqrt{\delta}$, and the associated eigenvalue is $\lambda (\delta)= \tilde{V}''(u_*) = 2 \sqrt{\delta}$. Now, assuming the Ansatz $\rho(\lambda) \propto \lambda^{\tilde\theta}$ for the left tail, we determine $\tilde\theta$ by requiring 
$$ \int_{0}^{\lambda(\delta)} \rho(\lambda') \dif \lambda' = 1/L^d \,, $$
giving $\tilde\theta = 1$, in agreement with Eq. \eqref{eq:linearDoS}. The prefactor $\sigma^{-2} r_f^2$ therein is expected to hold in the collective pinning regime, and can be understood by a dimensional argument: $\lambda$ has the same dimension as $V''(u) \sim \sigma / r_f$, so $\hat\lambda:=\lambda r_f / \sigma$ is dimensionless. We show in Fig.~\ref{fig:cumulativedistribution}(b) that the linear tail of $\hat{\lambda}$ has a coefficient independent of $\sigma$ in our simulations (where $r_f=1$). The above arguments show Eq. \eqref{eq:linearDoS} for depinning transition in all dimensions, explaining in particular the numerical observation in 1d. 

In terms of vibration frequencies, Eq. \eqref{eq:linearDoS} and the relation \eqref{eq:omega} give a universal spectrum $D(\omega) \propto \omega^3$ at low frequencies for the depinning models. Remark that this is qualitatively reminiscent of, yet quantitatively distinct from the $D(\omega)\sim \omega^2$ universal behavior found in densely packed spheres in high dimensions~\cite{charbonneau16universal}.

\subsection{Depinning soft modes}
\begin{figure}[h]
(a)\includegraphics[width=.97\columnwidth,valign=t]{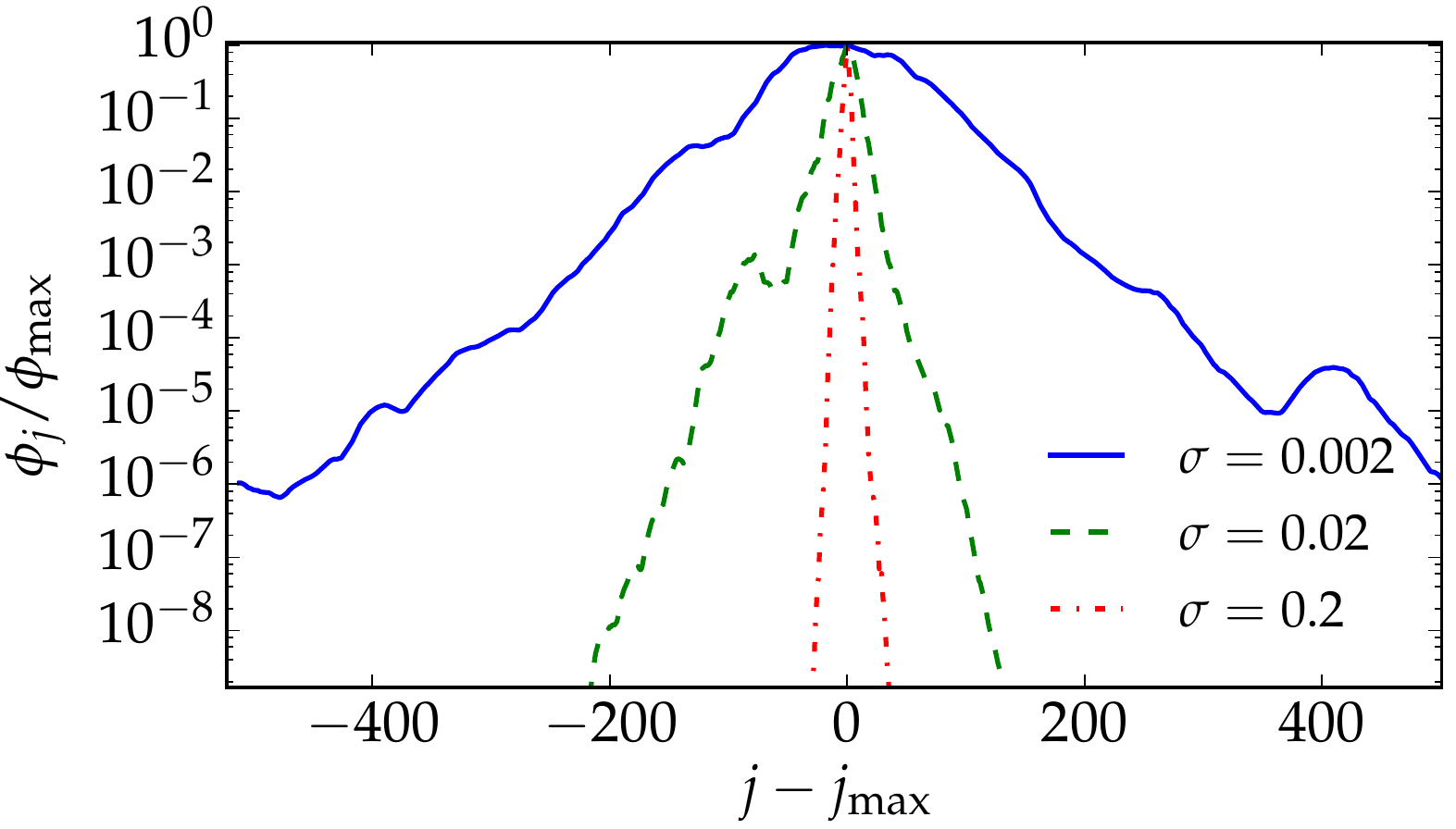} 
(b)\includegraphics[width=.97\columnwidth,valign=t]{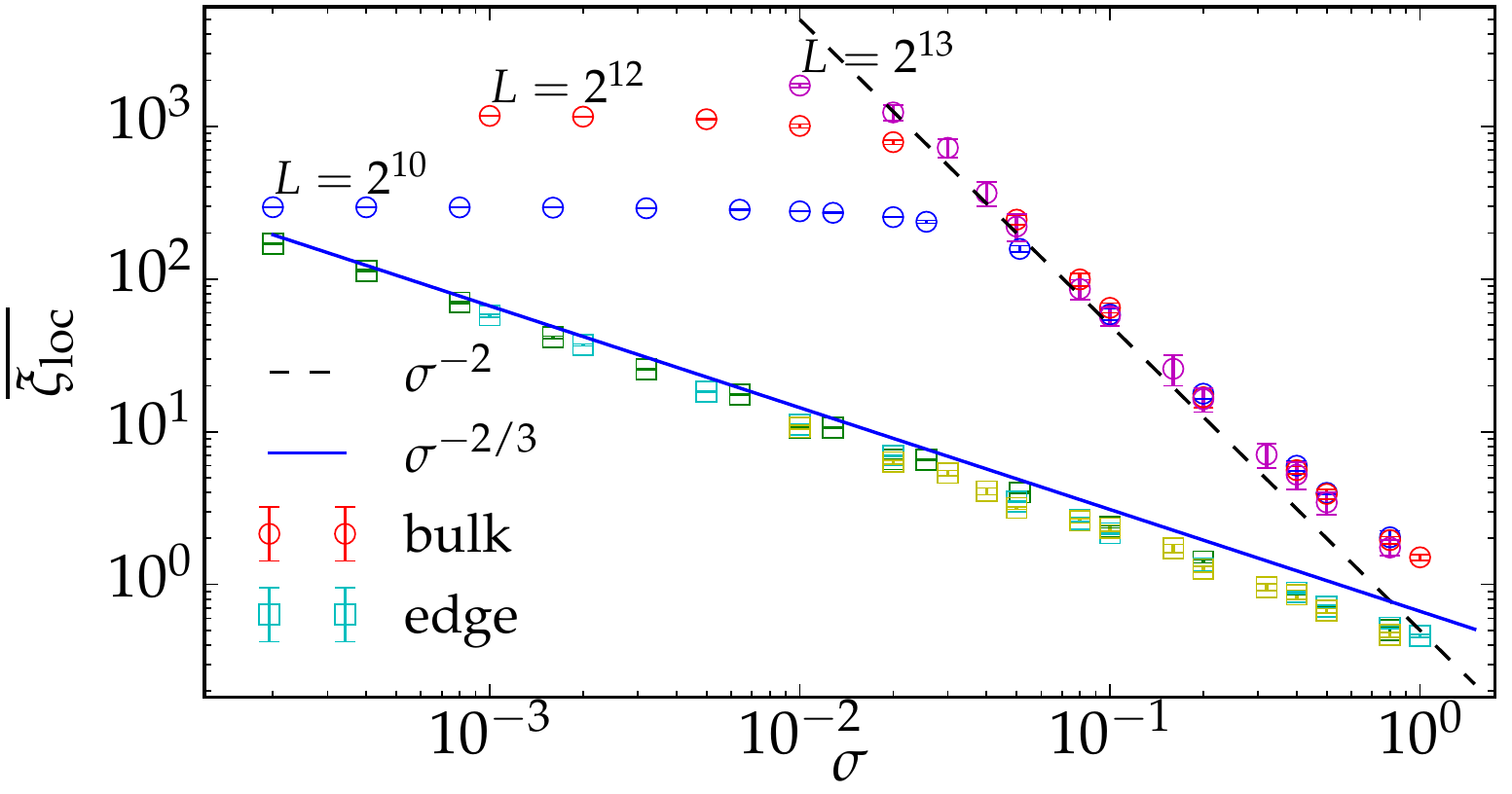}
\caption{(a)  
Samples of dynamical matrix ground states of 1D short-range depinning model of $L = 1024$, with $\sigma = 0.002, 0.02, 0.2$, centered around its maximum. (b) Disorder Average of the localization length ${\xi}_{\text{loc}}$ [eq.~\eqref{eq:xidef}] as a function of $\sigma$, of the ground state (``edge'') and of the 100 states in the middle of the spectrum (``bulk''). 
\label{fig:eigenvectors}}
\end{figure}

Despite the qualitative difference between the DoS we show here that the ground states of the two models are remarkably similar. Let us first characterize the depinning ground state $\phi_j$, which are the epicenter of the avalanche instability. In Fig.~\ref{fig:eigenvectors} (a) we show a few samples of $\phi_j$, for different disorder strengths. They are all localized around a well defined center, but the localization length $\xi_{\text{loc}}$ varies with $\sigma$. A practical way to define $\xi_{\text{loc}}$ is:
\begin{equation}  \label{eq:xidef}\xi_{\text{loc}}^2 = \sum_j  j^2 \abs{\phi_j}^2 - \left( \sum_j  j \abs{\phi_j}^2 \right)^2 \end{equation}
where the ground state is normalized as $\sum_j \abs{\phi_j}^2 = 1$. In Fig.~\ref{fig:eigenvectors} (b), we found that $\overline{\xi_{\text{loc}}} \sim \sigma^{-2/3}$, which is the same behavior as the Larkin length $\ell_c$, Eq. \eqref{eq:Larkinlenght}. Note that this behavior is characteristic of the lowest eigenstates, while in the middle of the spectrum, we found a different exponent $\overline{\xi_{\text{loc}}} \sim \sigma^{-2}$, which is also known to describe $\xi_{\text{loc}}$ of the 1d Anderson model in the middle of the spectrum. 

\begin{figure}[h]
(a)\includegraphics[width=.95\columnwidth,valign=t]{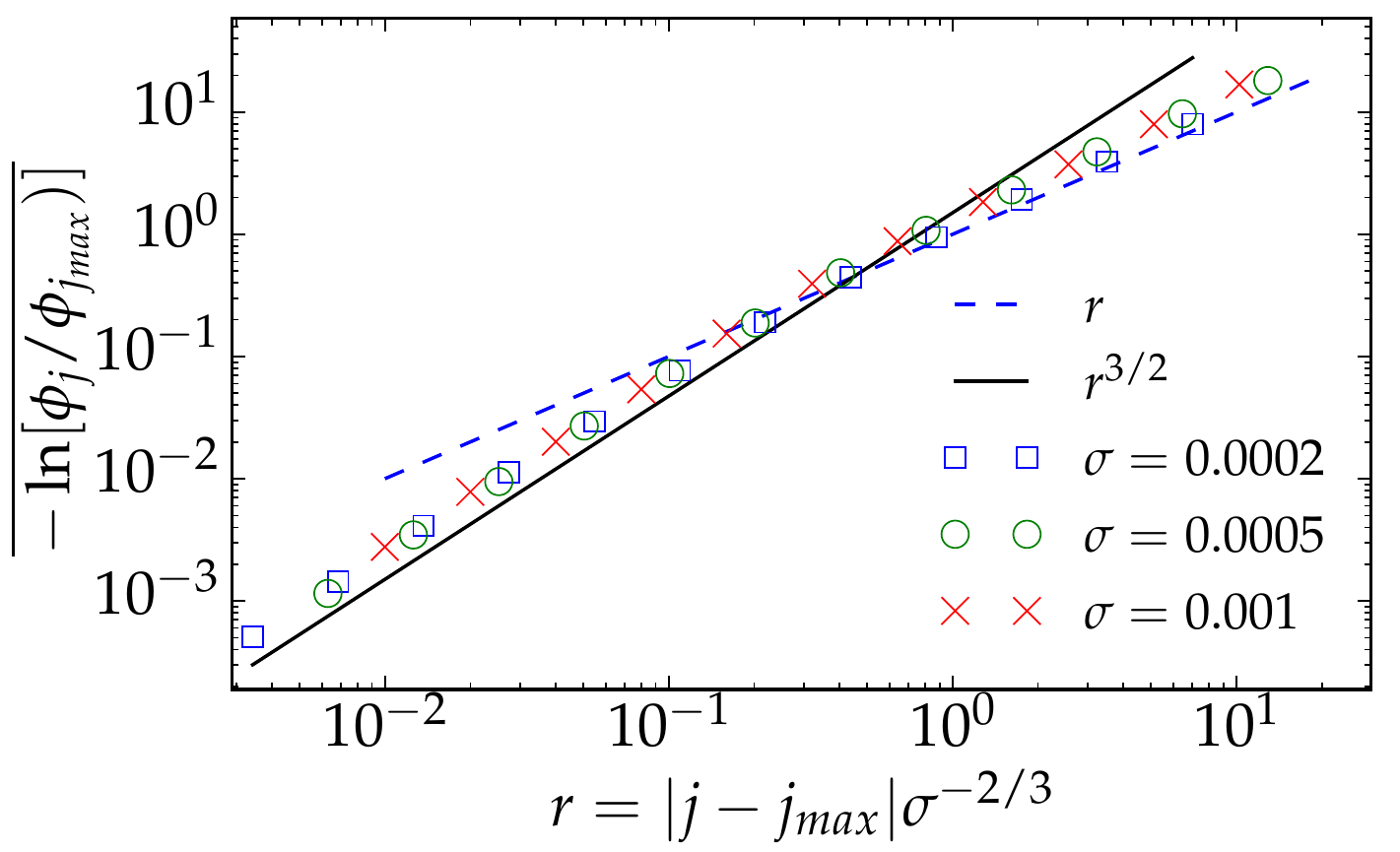}
(b)\includegraphics[width=.95\columnwidth,valign=t]{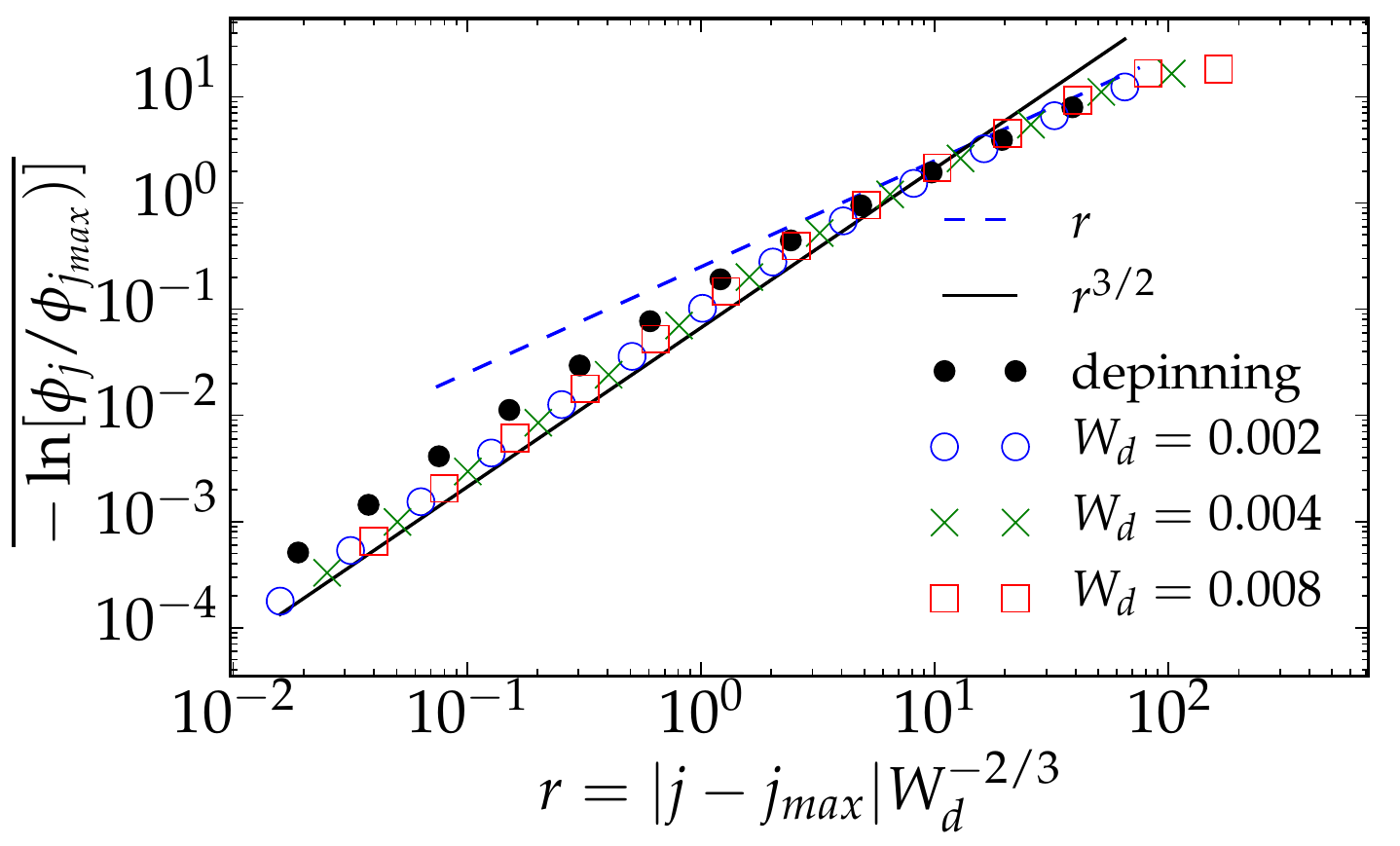}
\caption{(a) Numerical measure of the stretched-exponential decay of the ground state of the dynamical matrix, of size $L = 2^{12}$, and of different disorder strengths. They compare well to the Ansatz Eq. \eqref{eq:conjecture}. (b): The same measure for the short-range, 1D Anderson model. The matrices have sizes $L = 2^{13}$. The diagonal disorders are drawn from a uniform distribution in $[0, W_d]$. The ``depinning'' data is that with $\sigma = 0.0002$, which matches numerically to $W_d = 0.0026$ with Eq. \eqref{eq:Wdmatch}. The comparison shows a discrepancy in the stretched-exponential prefactor between Anderson and dynamical matrix ensembles.
}\label{fig:andersondepinning}
\end{figure}

More can be said about the shape of the ground state around its localization center $j_{\max}$, defined as the site where $\phi_j$ is maximum (recall that $\phi_j$ does not change sign, so we set $\phi_j \geq 0$). As we see in Fig.~\ref{fig:andersondepinning} (a), the decay of $\phi_j$ has two regimes: 
\begin{equation}
 \overline{- \ln (\phi_j/\phi_{\max})} \sim \begin{cases}
 \abs{(j-j_{\max})/\ell_c}^{\frac32}  \,,\, \abs{j-j_{\max}} \ll \ell_c  \\
 \abs{(j-j_{\max})/\ell_c}  \,,\, \abs{j-j_{\max}} \gg  \ell_c
\end{cases}\label{eq:conjecture}
\end{equation} 
where the characteristic length scale $\ell_c \sim \sigma^{-2/3}$ is again proportional to the Larkin length. It is interesting now to compare with the ground state of the Anderson model, with the protocol defined in Eq. \eqref{eq:Wdmatch}. Our results, in Fig.~\ref{fig:andersondepinning}, display the same scaling form as Eq.~\eqref{eq:conjecture}. However, we observe that the pre-factor in front of the $\abs{\frac{j-j_{\max}}{\ell_c}}^{\frac32}$ term is larger for the depinning. We have no clear explanation for this discrepancy: in particular, the negative correlation between the diagonal elements of the dynamical matrix [see Fig.~\ref{fig:diagonal} (b)] would give a smaller pre-factor. 

While the exponential far tail of the Anderson model's ground state is expected from rigorous results~\cite{Klopp2002}, the stretched-exponential behavior and the identification between the ground-state localization length and the depinning Larkin length are discussed in the appendix~\ref{sec:riccati} and extended to generic $D<D_{uc}$. 

\section{Fully connected depinning}\label{sec:MF}
\label{sec:mf}
\begin{figure*}
	\includegraphics[width=.8\textwidth]{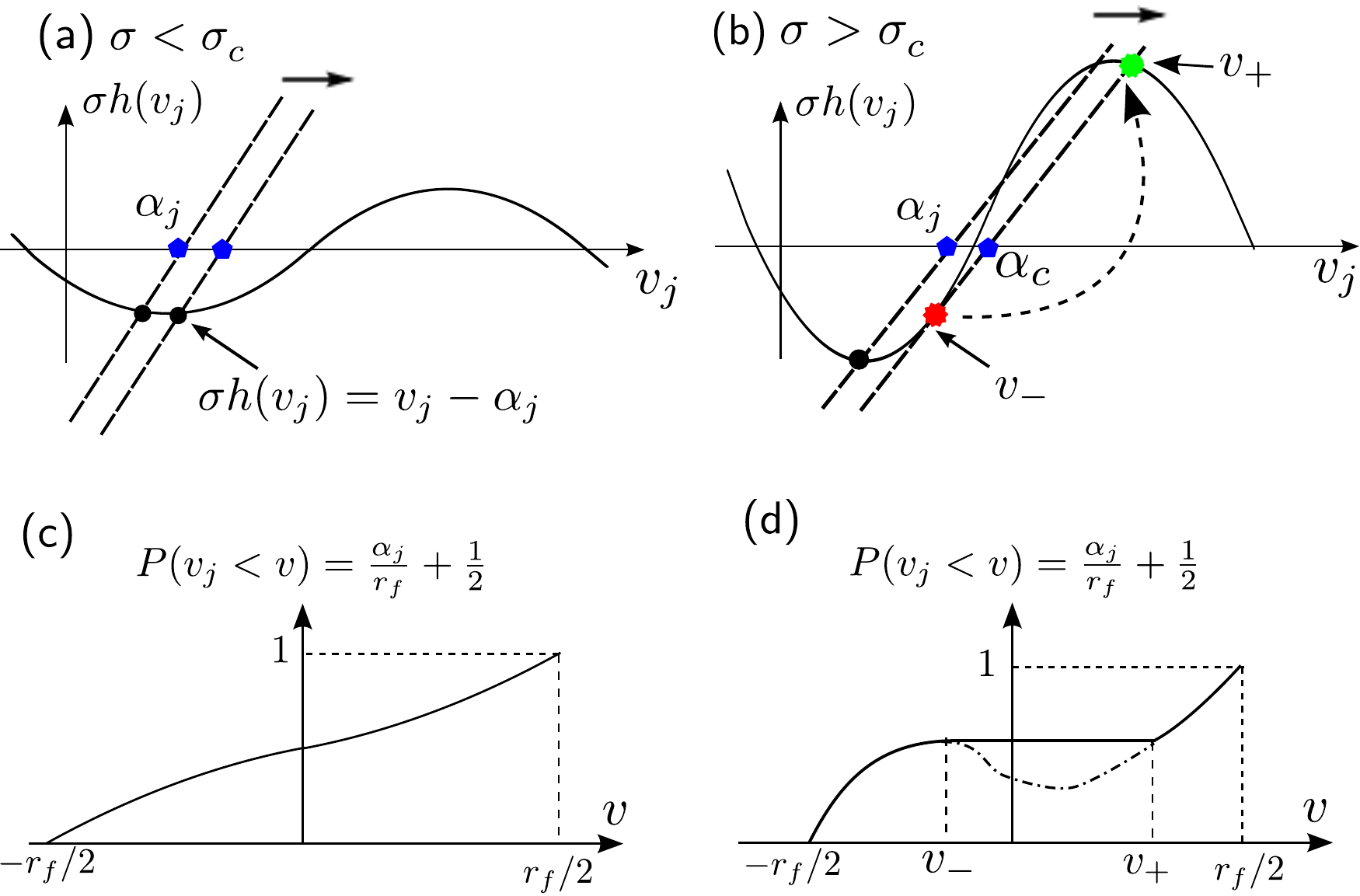}
	\caption{(a) Illustration of Eq. \eqref{eq:fuisuminusalpha} in the weak pinning phase; only one solution of $v_j$ exists for each $\alpha_j$. (b) Illustration of Eq. \eqref{eq:fuisuminusalpha} and Eq. \eqref{eq:shock} in the strong pinning phase. (c,d) The cumulative distribution of $v_j$ Eq. \eqref{eq:Pu} in the weak and strong pinning phases. In (d), the dash-dotted curve corresponds to the unstable solution of eq. \eqref{eq:fuisuminusalpha} that is discarded, see main text.
		}\label{fig:fullsketch}
\end{figure*} 
Up to now, the ground state of the dynamical matrix is always found to be localized, and corresponds to the epicenter of an avalanche.  In this section, we discuss a notable exception to this picture: the fully connected model with periodic force (with $\alpha = 0$ in Eq. \eqref{eq:frac}). This case was studied in Refs. \cite{fisher1983threshold,fisher85} because of its relevance for charge density waves, and can be exactly solved. It displays a transition between weak and strong pinning phases. In the following, we characterize the ground state of the dynamical matrix, and show that: it is delocalized in the weak pinning phase and displays an intriguing localization in the strong pinning phase: a finite fraction of the ground state condensates on a few sites, while the rest organizes in a multi-fractal way. The analytic treatment presented in section \ref{sec:fullyfundamental} will be followed by numerical studies in section~\ref{sec:fullynum}.																																					

The Equation of motion of the fully connected model can be as:
\begin{equation} \dot u_j = \bar{u} -u_j +f + \sigma h(u_j + \beta_j)  \,. \label{eqfisher} \end{equation}
Here, $\bar{u} = \frac1L \sum_{j=1}^L u_j$ is the mean position, $h$ is the force profile with period $r_f$, and normalized in the following way:
\begin{equation} \max_{u \in [0, r_f]} h'(u)=1  \,. \label{eq:maxhprime} \end{equation}
Note that the disorder strength $\sigma$ in Eq. \eqref{eqfisher} is proportional to the one in Eq. \eqref{eq:correlator}. $\sigma$ is also constant for all sites $j$, and the only randomness resides in the phases $\beta_j$, which are independent and uniformly distributed in $[0, r_f]$. , the s  

A pinned configuration corresponds to a stationary solution of Eq.(\ref{eqfisher}):
\begin{align}
 &  h(v_j) =(v_j - \alpha_j)/\sigma  \,,  \nonumber\\
\text{with } & \alpha_j = \bar{u} + \beta_j+f \,, \; v_j=u_j+\beta_j \,. \label{eq:fuisuminusalpha}
\end{align}
The solution of the fully connected model is obtained by considering the response of a single site $j$ to its environment. By periodicity, we can restrict the $\alpha_j$ and $v_j$ within the interval $(-r_f/2,r_f/2)$. When the driving force $f$ and/or $\bar{u}$ increases, so does $\alpha_j$. As shown in Fig.~\ref{fig:fullsketch}, two cases should be distinguished as a function of $\sigma$:
\begin{itemize}
\item if $\sigma \leq \sigma_c= 1$ a unique solution exists, so that $v_j$ changes smoothly when $\alpha_j$ is increased. 
\item if $\sigma >\sigma_c=  1$ multiple solutions can be found and the dynamics selects the smallest one. As a consequence the solution $v_j$ displays a shock jumping from $v_-$ to $v_+$ when $\alpha_j$ reaches the critical value $\alpha_c$. The shock is described by the following Equations: \begin{align}
 &    h'(v_-) = 1/\sigma \,, \;   h''(v_-)>0  \nonumber\\
 &   \alpha_c =   v_- - \sigma \,  h(v_-)   \,. \label{eq:shock}
\end{align}  
\end{itemize}
A remarkable property~\cite{fisher1983threshold} of the thermodynamic limit is that, in the stationary solution, the variables $\alpha_j$ become uniformly distributed in $(-r_f/2,r_f/2)$. This is a consequence of the uniform distribution of the random phases $\beta_j$, and it is checked numerically in Fig.~\ref{fig:alphaj}, both above and below $\sigma_c=1$. This property and Eq. \eqref{eq:fuisuminusalpha} determines the cumulative distribution of the variables $v_j$: 
\begin{equation} P(v_j <  v)=(v-\sigma\, h(v))/r_f + C  \,,\label{eq:Pu}\end{equation} 
where $C$ is an integral constant to make sure that $P(v_j < -r_f/2) = 0$. As illustrated in Fig.~\ref{fig:fullsketch}, Eq. \eqref{eq:Pu} is valid for all $v \in (-r_f / 2, r_f/2)$ in the weak pinning phase; while in the strong pinning phase, it is valid only for $v \notin (v_-, v_+)$; for $v \in (v_-, v_+)$, $P(v_j < v)$ is constant. 

These results allow to compute the critical force $f_c$, as we recall in appendix \ref{sec:fc>0}. In particular, in the thermodynamic limit, $f_c = 0$ when $\sigma \leq 1$: this is the weak pinning phase. In the strong pinning phase $\sigma > 1$, $f_c > 0$ and can be explicitly computed.

\subsection{Depinning soft mode : exact results}\label{sec:fullyfundamental}
Now let us come to the (de)-localization of the ground state of the dynamical matrix, which has the following form in the fully-connected case:
\begin{equation}
M_{jk} = \delta_{jk} W_j + \frac1L \,,\, W_j =  -\sigma h'(v_j) +  1 \,, \label{eq:Wj}
\end{equation}
Such matrices can be exactly diagonalized, see appendix \ref{sec:fullymath}. The main result is that, $M_{jk}$ is marginally stable if and only if 
\begin{equation}
\frac1L \sum_{j=0}^{L-1} \frac{1}{W_j} = 1 \,, \label{eq:marginalfully}
\end{equation}
in which case, the ground state is given by 
\begin{equation}
\phi_j = (L {W_j})^{-1} \Rightarrow \sum_{j=0}^{L-1} \phi_j = 1 \,. \label{eq:phij}
\end{equation}
 Since there is no spatial structure in the fully connected model, one may assume that $W_0 \leq W_1 \leq \dots$, so that $\phi_0 \geq \phi_1 \geq \dots$. 

Naively, the distribution of the diagonal coefficients $W_j$ is given by Eq. \eqref{eq:Wj} and Eq. \eqref{eq:Pu}:
\begin{align}
& p(W) \dif W =  \frac1{r_f} (1 - \sigma h'(v) ) \dif v \,, \label{eq:rhoW} 
\end{align}
where $W = 1 - \sigma h'(v)$.  We shall see that Eq. \eqref{eq:rhoW} is correct when $\sigma \leq 1$ but is only partially correct when $\sigma > 1$.

To see this, let us consider the marginally stability condition Eq. \eqref{eq:marginalfully}, assuming that $p(W)$ correctly describes the value distribution of $W_j$. In the $\sigma < 1$ phase, we have 
\begin{align}
&\frac1L \sum_{j=0}^{L-1} \frac{1}{W_j} = \int \frac1W p(W) \dif W  \nonumber \\ 
=  &\frac1{r_f} \int_0^{r_f} \frac1{1 - \sigma h'(v)} (1 - \sigma h'(v)) \dif v = 1 \,. \label{eq:marginalweak}
\end{align}
So, the ground state is always marginally stable in the $\sigma < 1$ phase. This should not be interpreted as the existence of avalanches, as $\phi$ is extended. Indeed, Eqs. \eqref{eq:phij} and Eq. \eqref{eq:maxhprime} imply immediately that 
$$\phi_{\max}  \leq \frac{1}{L (1 - \sigma)} \,.$$
Therefore, the ground state $\phi$ is extended \footnote{One can also show that $\phi_{\min}  \geq L^{-1} (1 - \min_v \sigma h'(v))^{-1}$; so the coefficients are uniformly bounded from below as well as above.}. For any given $h(v)$, the value distribution of the coefficients can be also exactly predicted by assuming the uniform distribution of $\alpha_j$.  

When $\sigma > 1$, the calculation analogous to Eq. \eqref{eq:marginalweak} would yield:
\begin{align}
&\frac1L \sum_{j=0}^{L-1} \frac{1}{W_j} \approx \int \frac1W p(W) \dif W \nonumber \\
=  &\frac1{r_f} \int_{v_+}^{r_f + v_-} \frac1{1 - \sigma h'(v)} (1 - \sigma h'(v)) \dif v \nonumber \\
 = & 1 -  n_0 < 1 \,,\,  n_0:= \frac{v_+ - v_-}{r_f} \,. \label{eq:notzero}
\end{align}
However, this does \textit{not} mean that, in the phase $\sigma > 1$, the lowest eigenvalue cannot vanish. Indeed, Eq. \eqref{eq:notzero} is a continuum calculation, which does not take into account correctly the contribution of a few smallest $W_j$'s, which are close to $0$. Their contribution is precisely $n_0$. This phenomenon is formally comparable to the Bose Einstein condensation (BEC), and $n_0$ is analogous to the macroscopic occupation number of the lowest one-particle mode. So, the ground state has two types of coefficients: the ``discrete'' ones, giving a total contribution $n_0$ ($\phi_{\max} \lesssim n_0$), and the ``continuum'' ones, whose largest coefficients are as follows: 
 \begin{align}
 \phi_j \approx \sqrt{\frac{c}{jL}} \,,
 \label{eq:phijdecay}
 \end{align}
 where $c =  1 / \sqrt{ 2 r_f \sigma h''(v_-) }$ is a constant.  

To derive Eq. \eqref{eq:phijdecay}, we observe that the largest $\phi_j$'s correspond to the smallest $W_j$'s, which in turn are associated with $v_j$'s closest to $v_-$, since 
$$ W_j = 1 - \sigma h'(v_j) \,. $$ 
Again, these $v_j$ are obtained by considering $\alpha_j$'s closest to the threshold value $\alpha_c$: 
$$  v_j - \sigma h(v_j) = \alpha_j \,.$$
By Equi-distribution of $\alpha$, we can write $\alpha_j = \alpha_c - j r_f / L$.

Now, we expand $h(v)$ around $v_-$, using Eq. \eqref{eq:shock}:
\begin{equation} \sigma h(v) =  -\alpha_c + v  +  \sigma h''(v_-) (v-v_-)^2 / 2 + \dots  \,. 
\label{eq:fuexpandstr} \end{equation} 
Using this expansion and the previous observations, we obtain 
\begin{equation}
W_j = \sigma  h''(v_-) (v_- - v_j) = \sqrt{2 \sigma h''(v_-) (\alpha_c - \alpha_j)} \,. \label{eq:Wjlim}
\end{equation}
Using this Equation and $\alpha_c - \alpha = j r_f/L$, we obtain Eq. \eqref{eq:phijdecay}. 

Note that Eq.~\eqref{eq:phijdecay} should not be interpreted as a power law tail, similar to the ones observed in Sect.~\ref{sec:LR} for long-range models.  because in the fully connected model, the notion of distance is trivial. Any two distinct points are far away from each other (in the sense of the $1/N$ elastic interaction). In this point of view, Eq. \eqref{eq:phijdecay}, for $j$ small, describes the amplitude of the largest secondary peaks among the continuum coefficients. 

In conclusion, the soft mode of the fully connected model is quite peculiar: a macroscopic fraction of the total mass is localized on a few sites, yet the rest has a multi-fractal structure. 

As an additional result, we consider the DoS of the dynamical matrix. For this we use Eq. \eqref{eq:Wjlim} to compute $p(W)$ near $W \sim 0$ for $\sigma > 1$, by recalling that $\alpha$ is uniformly distributed in $[-r_f/2, r_f/2]$: 
\begin{equation}
p(W) \sim \frac{1}{r_f \sigma h''(v_-)} W \,,\, 0 <  W \ll 1 \,. \label{eq:pW}
\end{equation}
By the interlacing relation Eq. \eqref{eq:interlacing} and the fact that $\lambda_0 = 0$, we deduce that the DoS of the dynamical matrix has the same linear left-tail $\rho(\lambda) = p(\lambda = W) \sim \lambda$, in agreement with Eq. \eqref{eq:linearDoS}.

\subsection{Numerical results : critical force and soft mode}\label{sec:fullynum}
 We verify numerically the key results in the section. In Ref.~\cite{fisher1983threshold,fisher85}, $h(u)$ is a sinusoid: $h(u) = \sin (u)$, so that $r_f =  2 \pi$. Our  simulations are performed with a numerically more convenient  piecewise parabolic potential:
\begin{equation}
h(u) = \begin{cases}
u(u+1) \,,\, & u \in (-1,0] \,, \\
u(u-1) \,,\, & u \in (0, 1] \,. \\
\end{cases} \label{eq:fquadratic}
\end{equation}
This function of period $r_f=2$ has continuous first derivative, but its second derivative diplays jumps when $u$ is an integer.

\begin{figure}[h]
	\includegraphics[width=1\columnwidth]{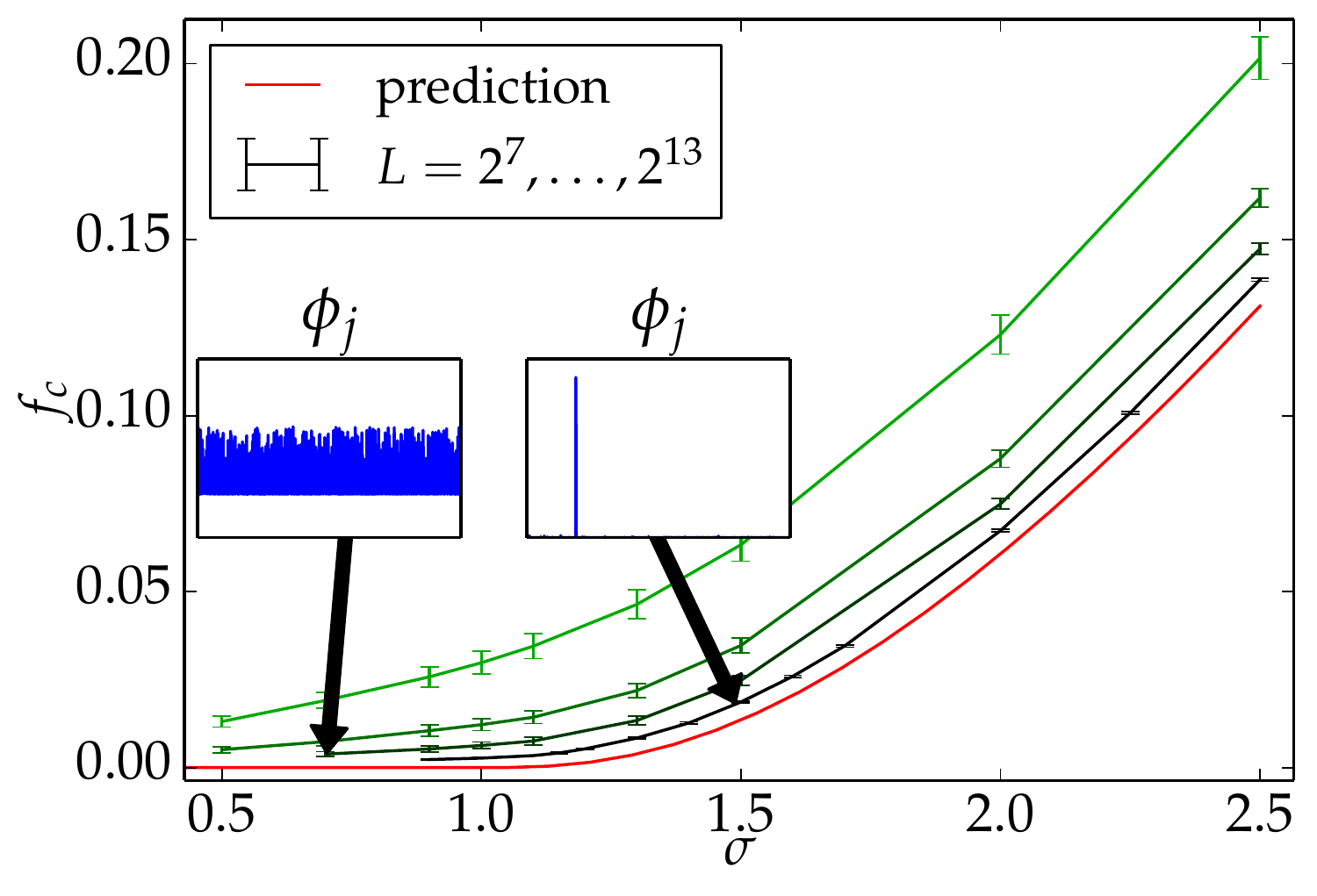}
	\caption{\textit{Main plot}: numerical measure of the critical force $f_c$ in the fully connected model, with the quadratic disorder Eq. \eqref{eq:fquadratic}, and system sizes $2^7, 2^9, 2^{11}, 2^{13}$ (from up to down). The theoretical prediction of $f_c$ is given by Eq. \eqref{eq:fcquadratic} in the $\sigma > 1$ phase, and vanishes when $\sigma \leq 1$. \textit{Insets}: the ground state of the pinned configuration, with $\sigma = 0.7$ (left) and $\sigma = 1.5$ (right), and with $L = 2^{13}.$ See also Figure \ref{fig:phij}. }\label{fig:mf_fc}
\end{figure}

First, we consider the critical force, see Figure \ref{fig:mf_fc}. The analytical prediction in the thermodynamic limit can calculated using the general formula of Ref.~\cite{fisher1983threshold,fisher85}; in our case, we have:
\begin{equation}
f_c = \frac{\left(2 \sqrt{2}+3\right) (\sigma -1)^3}{24 \sigma ^2} \,,\, \sigma > 1 \,, \label{eq:fcquadratic}
\end{equation}
see appendix \ref{sec:fc>0} for more details. Eq. \eqref{eq:fcquadratic} is compared to numerical measures in finite system sizes, using the same algorithm as the short-range study above~\cite{werner02rough} (except that apply the force directly without using the soft spring). We observe that the numerical data are compatible with the analytical prediction, despite pronounced finite size effects: the critical force $f_{c,L}$ in a system of size $L$ decreases with respect to $L$. In particular, $f_{c,L} > 0$ in the $\sigma < 1$ phase as well as in the $\sigma > 1$ phase. As a consequence, even when $\sigma < 1$, we can find meta-stable configurations in finite systems. Note also that the analytical theory presented above does not allow to predict $f_{c,L}$ for $L < \infty$. We leave this interesting question to future study. The ground state of the dynamical matrix display a sharp qualitative change between $\sigma <1$ (delocalized) and $\sigma > 1$ (localized), in agreement with the prediction of section \ref{sec:fullyfundamental}.

\begin{figure}[h]
	(a)\includegraphics[width=.97\columnwidth,valign=t]{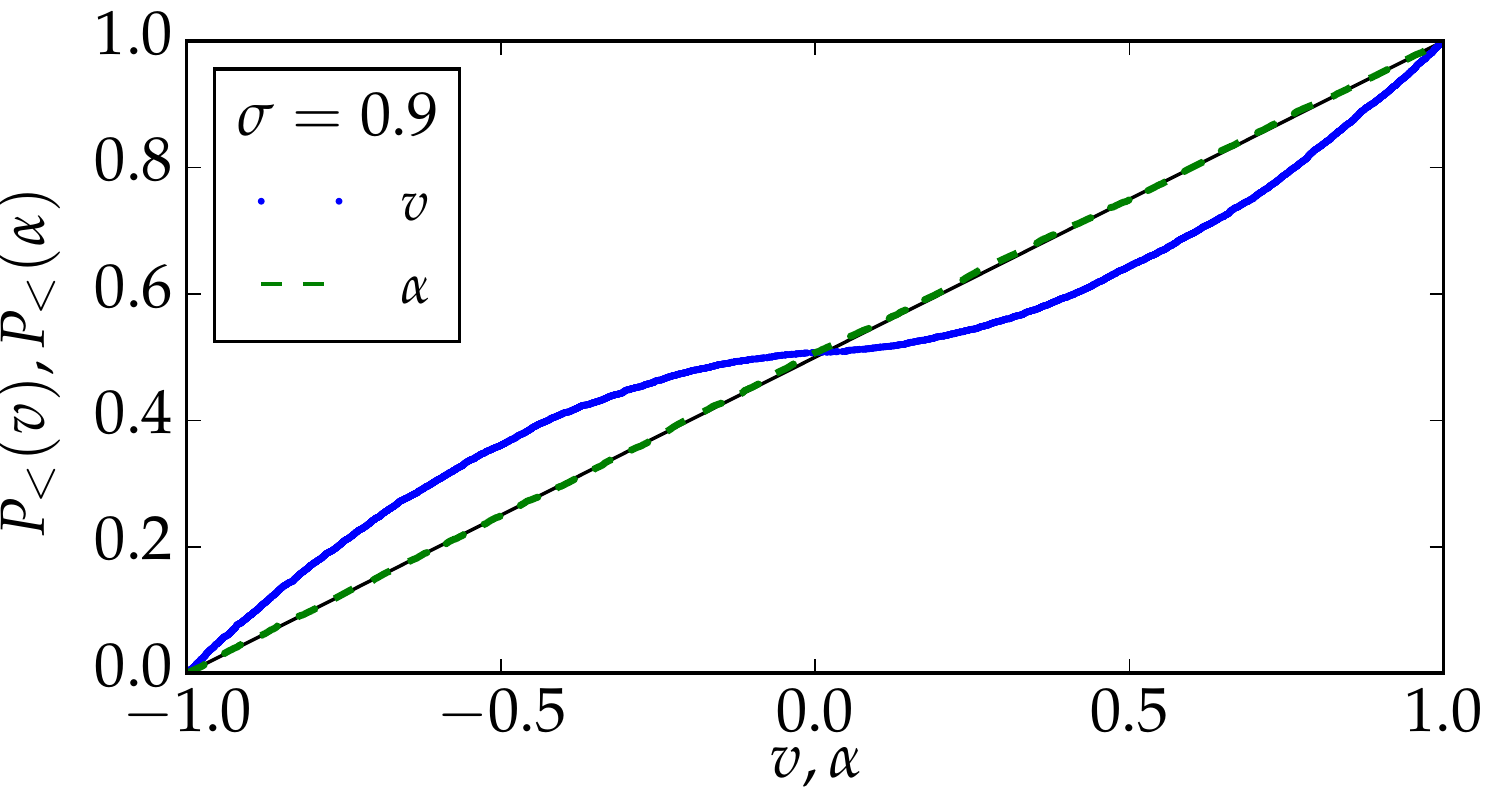}
	(b)\includegraphics[width=.97\columnwidth,valign=t]{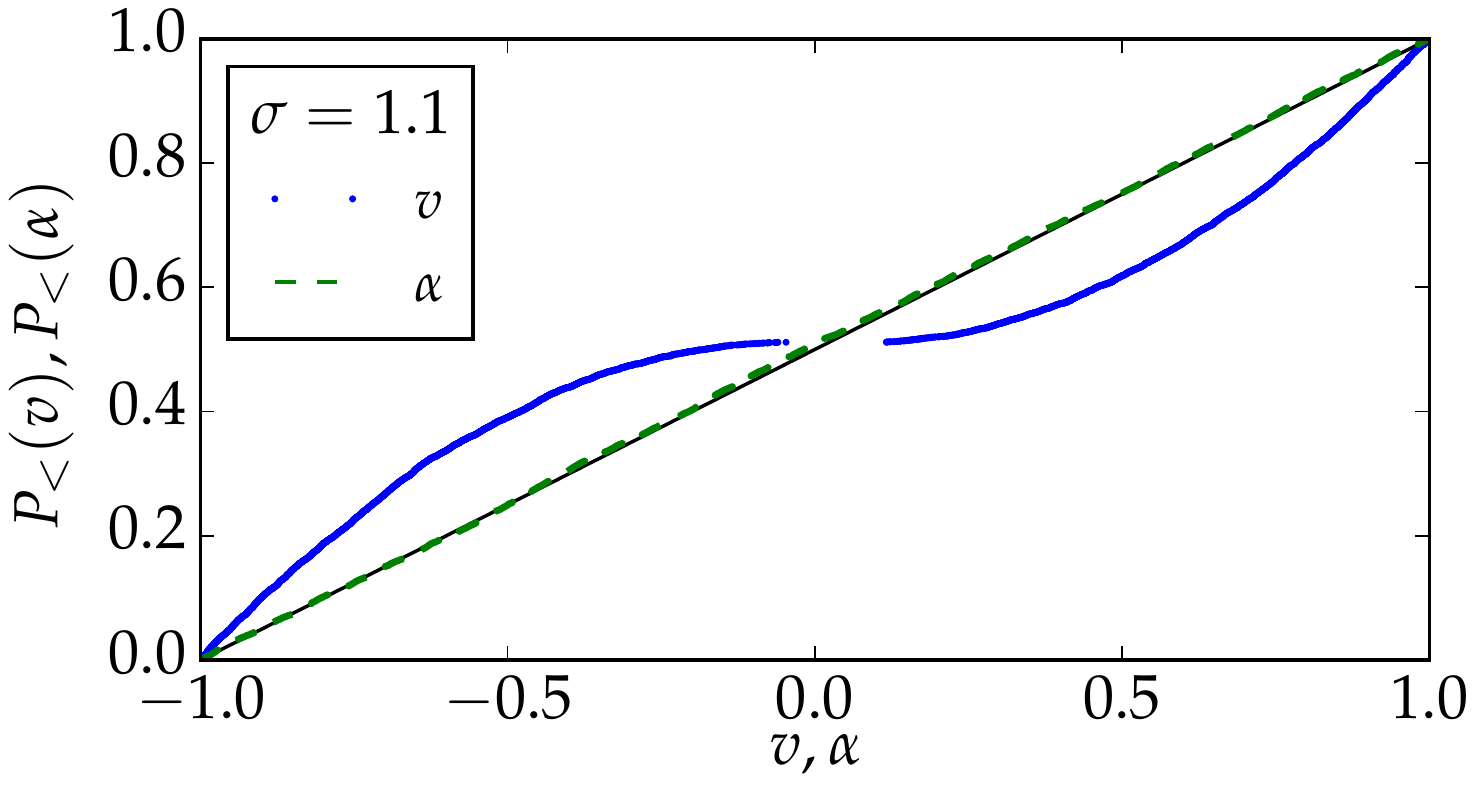}
	\caption{The cumulative distribution function of $\alpha_j$ and of $v_j$ ($P_<(\alpha)$ and $P_< (v)$, respectively), in two representative samples of $L = 2^{13}$, with $\sigma = 0.9 < 1$ (a) and $\sigma = 1.1 > 1$ (b), respectively. For both value of $\sigma$, the distribution of $\alpha$ agrees with the uniform distribution in the interval $(-1,1)$. When $\sigma=0.9<1$ the distribution of $v_j$ is continuous, while for $\sigma=1.1>1$, it displays a jump.}\label{fig:alphaj}
\end{figure}

Next, we verify directly the main assumption behind the analytical treatments presented: $\alpha_j$ are uniformly distributed. For this, we consider individual marginally stable configurations obtained in the measure of the critical force; no disorder averaging is involved. Knowing the random phases $\beta_j$ and the force profile, we calculate $v_j$ and $\alpha_j$ using Eq. \eqref{eq:fuisuminusalpha} (modulo the period $r_f = 2$), and then their cumulative distribution functions. The results obtained from representative samples in both phases are plotted in Figure \ref{fig:alphaj}. We observe that the distribution of $\alpha_j$ is always uniform, while  that of $v_j$ is non-uniform, and has a jump in the $\sigma > 1$ phase in particular. 
 
\begin{figure}[h]
	(a)\includegraphics[width=.97\columnwidth,valign=t]{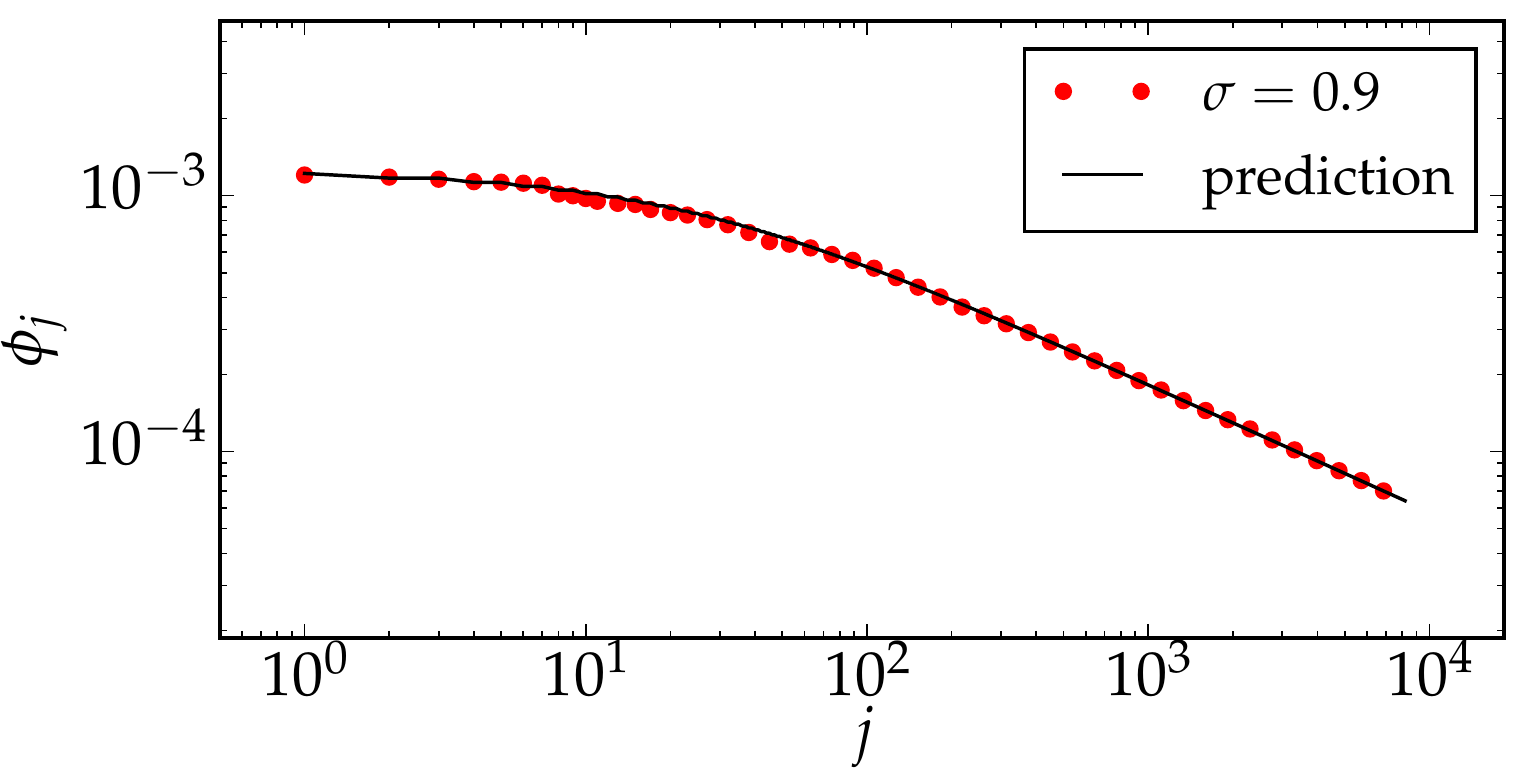}
    (b)\includegraphics[width=.97\columnwidth,valign=t]{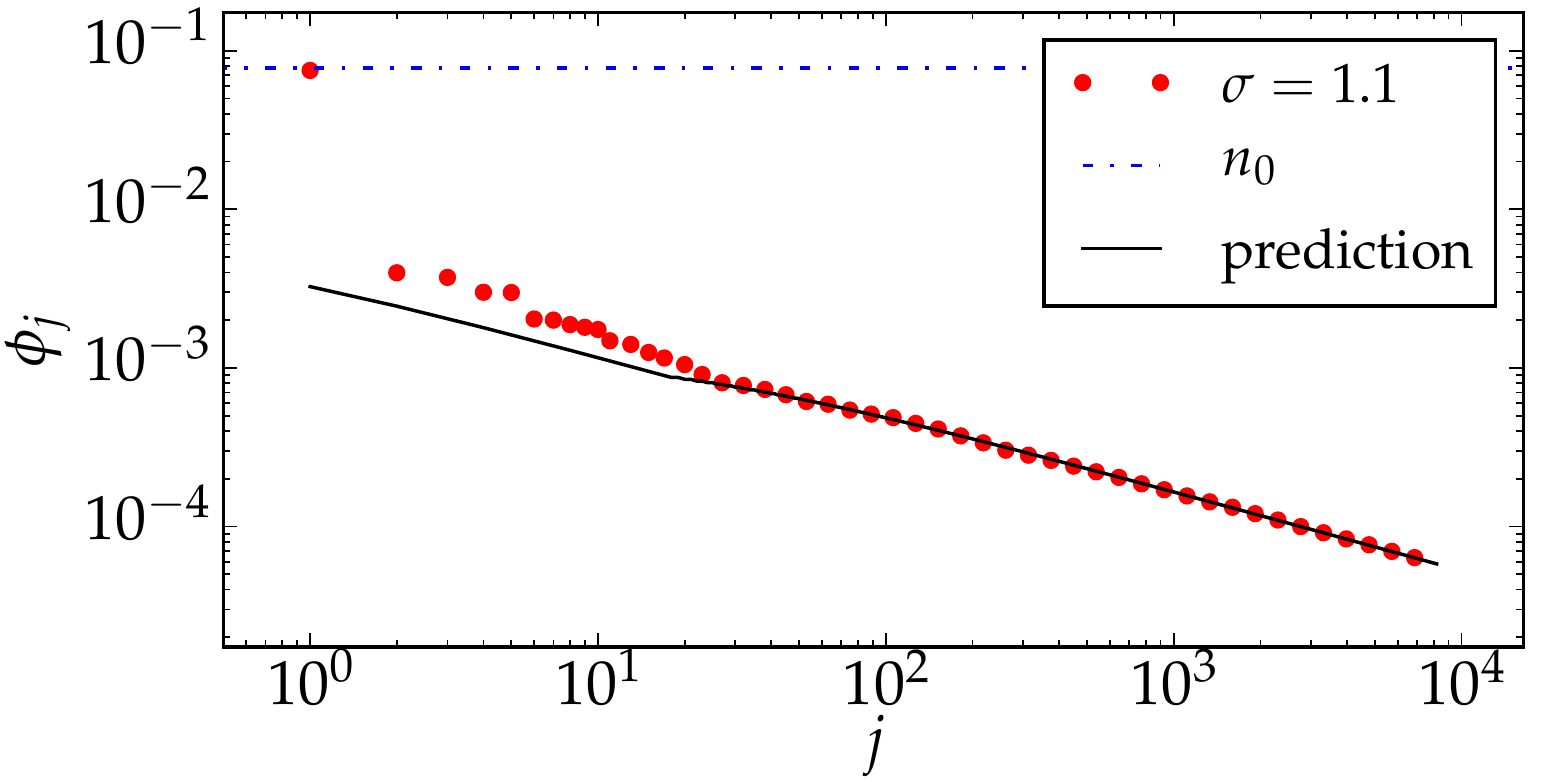}
\caption{The coefficients of the ground state, in the decreasing order. The analytical curve is obtained by Eq. \eqref{eq:Wj}, \eqref{eq:phij} by assuming that $\alpha_j = j U / L$ is Equi-distributed. In the $\sigma<1$ phase, it is expected to represent the whole ground state. In the $\sigma > 1$, it is expected to cover only the continuous part of the coefficients. }\label{fig:phij}
\end{figure}

Finally, we come to the ground state. Taking the same configurations as in Figure \ref{fig:alphaj}, we now calculate the coefficients of the ground state, by using the results of Appendix \ref{sec:fullymath}; the results are plotted in Figure \ref{fig:phij}. We observe that, in the weak pinning phase, all the coefficients are correctly described by the analytical prediction (of the continuous part), whereas in the strong pinning phase, there are clearly two kinds of coefficients: a few largest coefficients are much larger than the continuous prediction, while the rest is still in agreement with it. We also observe that, the fraction $n_0$ in Eq. \eqref{eq:notzero} is almost contributed by the single largest coefficient. This is in contrast with the short-range case (or more generally, $\alpha > d/2$ cases), where the ground state has a localization length. 
 
\section{Ground state localization of long-range Anderson models ($D > 4$)}\label{sec:LR}
In sec. \ref{sec:SR} we have seen that below $d=4$ for short range elasticity and in general below $D_{uc}$, soft modes are always localized. At weak disorder, when the pinning becomes collective, their localization length coincides with the Larkin length $\ell_c$. It is natural to ask what happens  above $D_{uc}=4$, where the elastic interface is flat, and $\ell_c \to \infty$. In sec.~\ref{sec:MF} the solution of fully connected ($D=\infty$) model shows the existence of two distinct phases by varying the disorder strength:
\begin{itemize}
 \item for $\sigma<\sigma_c$  the dynamics is characterized by the absence of pinnning ($f_c=0$) in the thermodynamic limit and by delocalized soft modes  
\item for $\sigma>\sigma_c$   the critical force is finite and  soft modes display atomistic localization and multi-fractal structure. 
\end{itemize}
We indentify the transition from weak to strong disorder with the delocalization-localization transition for the soft mode of the depinning dynamical matrix. In order to have a complete picture of the dynamical phase diagram of Fig. \ref{fig:interface} we should understand if long-range depinning models with $D>D_{uc}$ have a delocalized soft mode at small disorder.

 Unfortunately this is a very difficult task as depinning models are not integrable at finite $D$ and only moderate system sizes ($L\le 10^4$) can be achieved by direct numerical simulations. To make progress we rely on one of the main observations of section~\ref{sec:SR}: the statistical properties of the modes of the depinning dynamical matrix are very similar to those of the eigenvectors of the Anderson model. Assuming that such is still the case in the long range cases, it is then instructive to study the ground state properties of the generalized Anderson models defined in  Eq.~\eqref{eq:rodriguez}. We focus on this proxy problem in this section.
 
  Indeed, these models were studied by Rodriguez \textit{et. al.}~\cite{rodriguez03anderson}, motivated by other considerations. They showed that for $D>D_{uc}$ and small disorder part of the spectrum of the long-range Hamiltonian is delocalized.  However, at variance with the $d=3$ Anderson model, delocalization here occurs close to the edge and not at the center of the spectrum. Following their argument one could think that even the ground state delocalizes for $D > 4$ and small enough disorder.  Let us review their basic argument leading to that claim. 

In absence of the disorder ($W_j = 0$), the ground state of Eq. \eqref{eq:rodriguez} is a constant $\phi_j = L^{-d/2}$, and the first excited states are plane wave of wave-vector of length $1/L$; in 1d, it is written as $\psi_j = L^{-1/2} \exp(-2 \pi \im j / L)$. By Eq. \eqref{eq:frac}, the gap between their eigenvalues is 
\begin{equation}
\delta E = \lambda_1 - \lambda_0 \propto L^{-\alpha} \,.
\end{equation}
Now let us turn on a weak and uniform diagonal disorder with $W_j \in [0, W_d]$. Consider its effect by perturbation theory. For that we calculate the matrix element 
\begin{equation}
V := \left< \phi \vert W \vert \psi \right> \Rightarrow \abs{V} \sim L^{-d/2} \sigma / r_f \,,
\end{equation}
by central limit theorem. When $\alpha < \frac12$, $\delta E \gg \abs{V}$, and we expect that the ground state is weakly (as $L\to \infty$) perturbed, and remains almost flat. Using perturbation theory at the lowest order one expects the energy of this state is $\sim W_d/2$.
Ref. \cite{rodriguez03anderson} and follow-up works \cite{malyshev04,balagurov05phase,demoura05} supported the above reasoning by a few other methods: a renormalization group argument based on super-symmetry field theory, coherent potential approximation, numerical simulations of system size up to $L^d = 10^5$.  In particular  the method of Ref. \cite{balagurov05phase}, section III.B yields a better estimation of the energy correction, namely $W_d / 2- c W_d^{2}$, where $c$ is a constant. These results prove that in long range models, delocalization occurs close to the band edge, but they do not imply that the ground state of the pure model is still the ground of the weakly disorder model.
Following Lifshitz ideas we argue that the ground state of the disorder model is actually always localized and the results of our extensive numerical simulations support this physical intuition.

\subsection{Lifshitz argument}
We begin by considering the short range case for which exact results are known. For this, we observe the ideas~ \cite{rodriguez03anderson, malyshev04,balagurov05phase,demoura05} could have been applied to the short-range Anderson model with $d > 4$, and led to conclude that its ground state would be delocalized for small enough disorder. However, this claim is in contradiction with well-established facts: since Lifshitz~\cite{Lifshitz}, it is well-known that the lowest lying states of the short-range Anderson model in any dimension are always localized around one of the deepest valleys of the potential; this fact was rigorously established in Ref.~\cite{Klopp2002}. The arguments of Ref.~\cite{rodriguez03anderson} fail in this context, mainly because the lowest band edge (Lifshitz tail) is always in the strong disorder regime, to which methods of weak localization (perturbation theory, super-symmetric field theory) do not apply. Instead, the effect of on-site disorder is non-perturbative. Even when its amplitude is small, the lowest lying states are not extended, but localized around the deepest valleys of the potential  (it is quite common that rare events play a decisive rôle on localization properties of long-range hopping models, see~\cite{cao16loc} for a recent example). 

To make this point clear and explicit, we recall the argument of Lifshitz~\cite{Lifshitz} and extend it to the long range case of our interest here. For this, we note that the meaning of ``deep valley'' depends on the distribution of the on-site disorder energy.  For the uniform distribution it refers to a box of linear size $\ell$ neighboring sites on which the disorder is small: $W_j \in [0, \epsilon]$. Such valleys are rare and appear with probability $P = (\epsilon/W_d)^{\ell^d}$. Now, a wave function that is confined in such a box would have kinetic energy $E_K \sim \ell^{-\alpha}$ and potential energy $E_P \sim \epsilon$. Setting $\ell = \epsilon^{-1/\alpha}$ (so that $E_K \sim E_P$), we deduce a bound for the ground state energy $ \leq 2 \epsilon$ which holds almost always, as long as $ L > \exp\left( - \epsilon^{d/\alpha} \ln (W_d/\epsilon)  \right).$ Note that states of such low energy, $2 \epsilon$, are well below the ones obtained by perturbing the pure system, namely $W_d / 2- c W_d^{2}$.

The above argument indicates that, the ground state is sensitive to rare deep valleys, which are only present in large enough systems; otherwise, the ground state would appear delocalized. This gives rise to a transient delocalization behavior, which could be (wrongly) interpreted as a transition. We suspect this to be case of the numerical studies in Ref.~\cite{rodriguez03anderson} (and follow-up papers), which were done with system of moderate sizes, $L \leq 10^5$.  Here, we shall use a new iterative scheme  to find the ground state of the long-range Anderson model, which allows us to consider matrix sizes up to $L = 2^{24} \geq 10^7$, and to observe the predicted transient. 

\begin{figure}
	\includegraphics[width=1.\columnwidth]{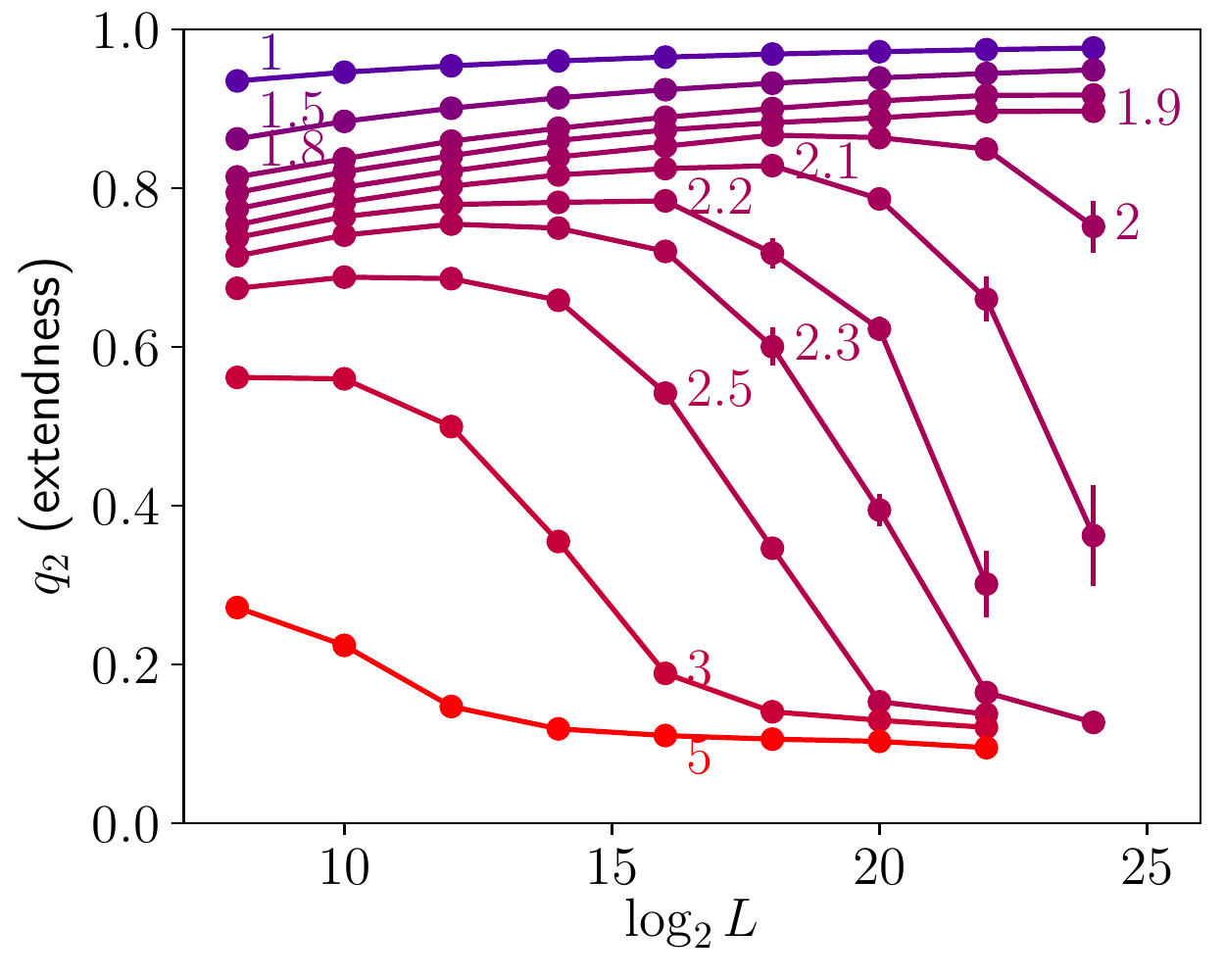}
	\caption{Localization of ground state in the 1d long range Anderson model, with $\alpha = 0.25$. The diagonal disorder $W_j$ are independent and uniformly distributed in $[0, W_d]$, where $W_d$ is indicated beside each curve using the same color. The inverse participation ratio exponent $q_{2}$, Eq. \eqref{eq:qinfty}, is averaged over $10^2 \sim 10^3$ realizations for each system size $L = 2^{8}, \dots, 2^{24}$. } \label{fig:LR}
\end{figure}

\subsection{Numerical set up}
 Let us first describe the iterative scheme. For this we denote $g = -G_{jj}$. We take the initial state to be uniform: $\phi_j^{(0)} = 1/L$, $j = 1,\dots, L$. The iterative step is given by:
\begin{subequations}\label{eq:algo}
\begin{align}
&\chi_j^{(n)} = \sum_{j \neq k} G_{jk} \phi_k^{(n)} \label{eq:step1} \\
&\phi_j^{(n+1)} = \frac{ \chi_j^{(n)} }{g  + W_j - E_n}  \label{eq:step3}
\end{align}
where $E_n$ is the unique solution to the following Equation: 
\begin{equation}
 \sum_{j = 1}^{L}  \frac{ \chi_j^{(n)} }{g  + W_j - E_n} = 1 \,, E_n < \min_j (W_j) + g \label{eq:step2}
\end{equation} 
\end{subequations}
In practice, given the vector $\left(\phi_j^{(n)}\right)$, we first calculate $\chi_j^{(n)}$ by Eq. \eqref{eq:step1}; thanks to the translation invariance of the elastic matrix $G_{jk}$, this can be done by fast Fourier transform. Then we solve Eq. \eqref{eq:step2} in the interval $E_n \in (-\infty, \min_j (W_j) + g)$ (by standard bisection or Newton routine), and obtain the new state $\phi_j^{(n+1)}$ by Eq. \eqref{eq:step3}. We stop the iteration when $E_n - E_{n-1}$ is smaller than a numerical tolerance (set to $10^{-8}$ in practice), and return the last $\phi_j$ as an accurate approximation of the ground state. 

For any long-range elastic matrix $G_{jk}$, the above iterative scheme has the following properties: 
\begin{itemize}
\item[-] If $\phi_j^{(n)} = \phi_j^{(n+1)} = \phi_j$ is a fixed point of the iteration, then it is a eigenstate of $M_{jk} = G_{jk} + \delta_{jk} W_j$ with energy $E(n)$. This follows immediately from Eq. \eqref{eq:step1} and \eqref{eq:step3}.
\item[-] At any step, the evolving states are positive and normalized as follows:
\begin{equation} \phi_j^{(n)} \geq 0 \,,\, \chi_j^{(n)} \geq 0 \,,\, \sum_{j = 1}^L \phi_j^{(n)} = 1 \,,
\end{equation} 
as one can show inductively using Eqs. \eqref{eq:algo}. Since the ground state is the only eigenstate of $M_{jk}$ with all-positive coefficients, the iteration is guaranteed to converge to the ground state. 
\item[-] For the fully-connected case, where $G_{jk} = 1/L$ for any $j \neq k$, the iteration converges exactly with one step: $\phi_j^{(1)} = \phi_j^{(2)} = \dots$ is the ground state. In this sense, our algorithm is a perturbation from the fully-connected case into $\alpha > 0$. Indeed, we observe numerically that the convergence is much faster for small $\alpha$, which is our interest here.
\end{itemize}

With the above iteration scheme, we studied the ground state of the 1d long-range Anderson model, with $W_j \in [0, W_d]$ independent and uniformly distributed, for different values of $W_d$. We use a standard measure of localization, the following inverse participation ratio exponent:
\begin{equation}
q_2 := - \frac{1}{\ln L} \ln \left[\frac{\sum_{j=1}^L \phi_j^4}{\left(\sum_j \phi^2_j\right)^2} \right]  \,. \label{eq:qinfty}
\end{equation}
For a uniform state $\phi^2_j  = 1/L$, $q_2 = 1$, while for a localized state whose maximal coefficient is independent of the system size, $q_{2} \to 0$. Our results are shown in Figure~\ref{fig:LR}. We observe that, for a large range of disorder strength $W_d \in [2, 3]$, as the system size increases, the ground state tends to be delocalized in a transient regime (up to $L = 2^{10}$ for $W_d = 3$ and $2^{10}$ for $W_d = 2$). For larger sizes however, the trend is rapidly reversed towards localization. Such a transient behavior agrees with the Lifshitz-type reasoning above, and was not noticed before. Note that the cross-over length grows very fast as $W_d$ decreases, so that for $W_d = 1$, the localization will occur at a prohibitively large system size. 

\begin{figure}
(a) \includegraphics[width = .9\columnwidth, valign=t]{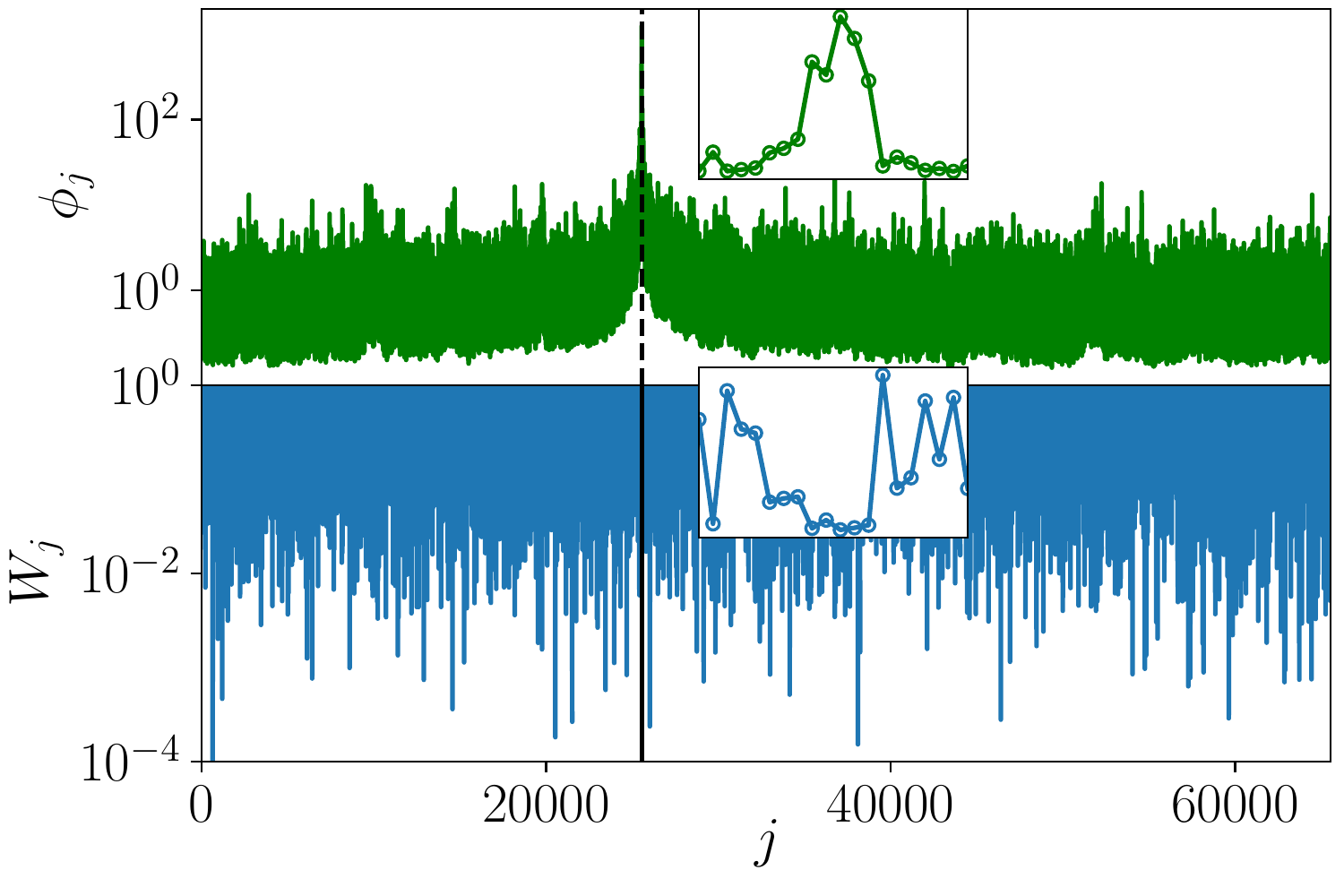}
(b) \includegraphics[width = .9\columnwidth, valign=t]{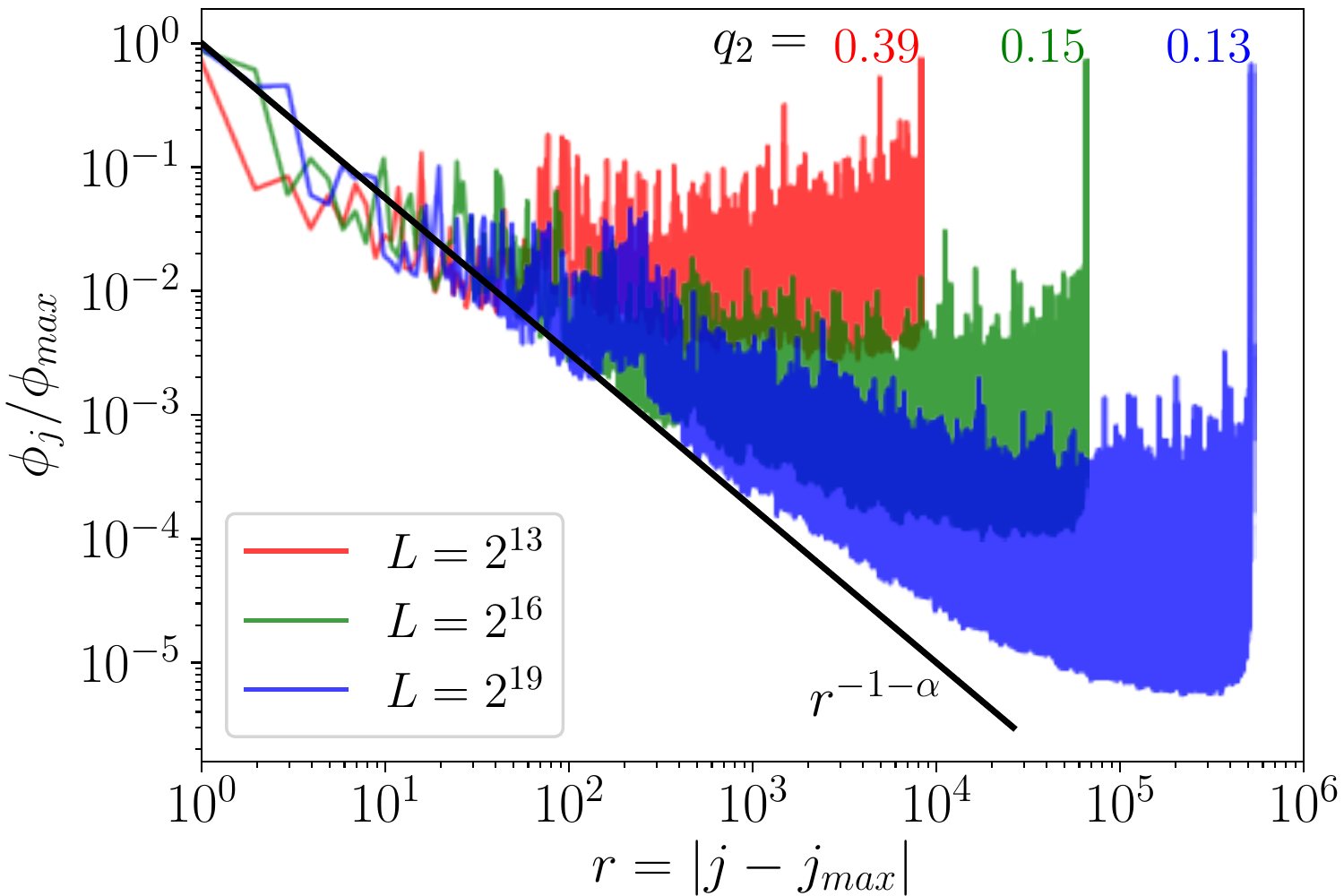}
\caption{(a) a sample of the disorder potential $W_j$ (in blue, lower curves, with $W_d = 3.0$) and the ground state of the corresponding long range Anderson model, with $\alpha = 0.25$. The system size is $L = 2^{16}$. The insets are zoom-in, linear scale plots of the disorder and the ground state near the maximal peak (indicated by black dashed lines in the main plot). (b) Ground states of $\alpha = 0.25$ long range Anderson model with $W_d = 3.0$ in different system sizes, with the maximum displaced to the origin. The wave-function decay is compare to the power law $r^{-\alpha - 1}$ argued in Ref.~\cite{balagurov05phase}. The value $q_2$, eq.~\eqref{eq:qinfty} of each wave-function is noted above it with the same color.}\label{fig:lrstate}
\end{figure}

The localized ground state in $D > 4$ long range Anderson models is peculiar compared to usual Lifshitz localization in their short range counterparts. Due to the power-law decay of the long-range hopping $G_{ij} \sim \abs{i-j}^{-1-\alpha}$, one expects a power-law decay of the wave-function away from its peak~\cite{balagurov05phase,Yeung87}: $\phi_j \propto r^{-1-\alpha}$, $r = \abs{j-j_{\max}}$. Upon numerical investigation of representative samples, see Fig.~\ref{fig:lrstate} (b), we observe indeed a power-law decay away from the localization center, and the associated exponent approaches from above the predicted value as the system size increases. We also observe many secondary peaks well visible in Fig.~\ref{fig:lrstate}, probably associated with other potential valleys. Note that such peaks are absent in the standard (short-range) Anderson model. Qualitatively, they are reminiscent of the multi-fractal structure of the continuum part of the localized soft mode in the fully connected depinning model. Quantitatively, in the fully connected case, the multi-fractality is not reflected in the inverse participation ratio exponent. Indeed we expect that $q_2 to 0$ as $L \to\infty$, due to the contribution of discrete coefficients; for instance, the representative ground state of Fig. 11 (b) has $q_2 \approx 0.01$. In the long range case, we observe numerically that $q_2$ seems to stagnate around $\sim 0.1$ rather than converging to $0$ in the system sizes accessible to us, see Fig.~\ref{fig:LR}. It is difficult to say whether it persists in the $L\to \infty$ limit and understand the origin of this behavior.

To conclude, we have shown, with an adapted Lifshitz argument and extensive numerical evidences, that the ground state of long range ($\alpha > 0$) Anderson models is never extended, but has always a localization peak around some rare deep potential valley. According to our general conjecture, for the depinning problem, we expect that the critical force is always finite in the thermodynamic limit when $D > 4, \alpha > 0$. 

\section{Conclusion}\label{sec:conclusion}
We revisited the disordered elastic model at the depinning transition, by focusing on the soft modes (lowest eigenstates) of the dynamical matrix of marginally stable configurations, which we compare to the generalized Anderson model with long-range hopping. We first show that the left-tail of the DoS has a linear behavior characteristic of all depinning models, at variance with the Lifshitz tails of the Anderson model. Yet, the ground state of the two random matrix ensembles are strikingly similar. In particular, we show that when the effective dimension $D = 2d / \alpha < D_{uc} = 4$, the localization length is identical to the Larkin length. Above $D_{uc}$, we argue that the localization always occur, in contradiction with the claim of Ref.~\cite{rodriguez03anderson}.

In the case of the fully connected depinning model, we showed that the ground state displays an intriguing localization in the strong pinning phase, and is extended in the weak pinning phase. This supports our general conjecture that: the soft modes of dynamical matrices of metastable configurations are localized and represents epicenters of avalanches; while in presence of weak pinning, the  ground state is always delocalized. 

We discuss some questions for future study. First and foremost, it is important to provide a first-principle understanding of the above localization-pinning conjecture. We believe that progress in this direction will help us resolve two other puzzles raised in this work: \texttt{(i)} how to rule out definitively the possibility of weak pinning in dimensions $ D_{uc} < D < \infty$; \texttt{(ii)} how to describe analytically the statistics of diagonal elements of the dynamical matrix and their effects on the ground state.

The depinning model considered here can be formally compared to the yielding model  \cite{Lin2014}, the main difference being that the elastic matrix $G_{ij}$ (the Eshelby kernel) has both positive and negative off-diagonal elements. This makes the ground state more challenging to study; in particular, its coefficient changes sign, and its localization properties are not known. 

It would be also interesting to explore the relevance of our results, particularly the dynamical matrix density of states left-tail, for understanding weak thermal effects at~\cite{middleton92rounding} or below the depinning threshold where a depinning-like coarse-grained dynamics is reported~\cite{Ferrero2017,Purrello2017}.

Recently, Ref. \cite{fyodorov2017exponential} proposed a new exciting  relation between Anderson localization and the depinning model, from the point of view of Equilibria counting. We remark that this viewpoint is heterogeneous to ours here, and it is a non-trivial task to connect them to each other.   

\begin{acknowledgments}
It is a pleasure to acknowledge discussions with M. Mueller, V. Démery, E. A. Jagla, P. Cornaglia, G. Usaj, D. Dominguez, V. Lecomte, P. Le Doussal, A. Nicolas, Ch. Texier, and D. Trimcev. X.C. acknowledges financial support from Capital Fund Management Paris and Laboratoire de Physique Théorique et Modèles Statistiques. This work is supported by PIP11220120100250CO (CONICET, Argentina), MINCYT-ECOS A16E01 (Argentina-France) and ANR-16-CE30-0023-01(THERMOLOC). 
\end{acknowledgments}


\appendix

\section{Fractional Laplacian in $d=1$}
\label{app:frac}
Here we give some detail on the fractional Laplacian defined on a  $d=1$  lattice. Extensions of the results to $d > 1$ are straightforward. 

Starting from the definition given in Eq. \eqref{eq:frac} the fractional Laplacian reads as:
\begin{equation}
G^{(\alpha)}_{ij} = -\frac{1}{L}\sum_{k=1}^{L-1}  e^{\im 2\pi k(i-j)/L} (2 - 2\cos(2\pi k / L))^{\alpha/2} \,, \label{eq:longrange}
\end{equation}
where $\alpha \in [0, 2]$. By construction, we have $\sum_{j=1}^L  G^{(\alpha)}_{ij} = 0$. Note that here periodic boundary conditions are implemented. Setting different boundary conditions (absorbing, reflecting,...) is a delicate issue discussed in Ref.\cite{zoia}.

For the intermediate values $\alpha \in (0,2)$, some exact expression can be derived when  $L \to \infty$ in  Eq. \eqref{eq:longrange}. Then, the discrete sum becomes a continuous integral, $2 \pi k / L \to \theta\in [0, 2\pi)$:
\begin{align}
G^{(\alpha)}_{ij} \to & -\int_0^{2\pi}  \frac{\dif \theta}{2\pi}  
e^{\im \theta (i-j)} (2 - 2\cos\theta)^{\alpha/2} \nonumber \\
= & \frac{\Gamma(\abs{i-j} - \alpha/2)\Gamma(\alpha+1) \sin(\alpha\pi/2)}{\pi \Gamma(\abs{i-j}+\alpha/2+1)} \,,
\end{align}
see \cite{zoia}, Eq. 8. Using Stirling formula for the Gamma functions, we deduce the asymptotic behavior as $ 1 \ll \abs{i-j} \ll L$:
\begin{align}
&  G^{(\alpha)}_{ij} \sim c_{\alpha} \abs{i-j}^{-\alpha - 1} \,,\, \alpha \in (0,2) \,. \label{eq:powerlaw} \\
&  c_\alpha = \Gamma(\alpha + 1) \sin (\alpha \pi / 2) / \pi \,.
\end{align} 
That is, the off-diagonal elements are non-positive: $G^{(\alpha)}_{ij} < 0$ when $i \neq j$ and their absolute value decays as a power of the distance. The sum of the absolute value of the off-diagonal elements on any row is Equal to 
$$G^{(\alpha)}_{ii} \stackrel{L\to \infty}\longrightarrow - \frac{2^{\alpha } \Gamma \left(\frac{\alpha +1}{2}\right)}{\sqrt{\pi } \Gamma \left(\frac{\alpha }{2}+1\right)} \,.$$

As a further example, we consider the de Gennes model, which corresponds to $\alpha = 1$. In this case, the above formulas simplify to the following explicit form:
\begin{equation*}
G^{(\alpha)}_{ij} \stackrel{L\to \infty}\longrightarrow \frac{4}{\pi\left(4 \abs{i-j} - 1\right)^2}\,.
\end{equation*}

In general, we can interpret long range interaction as mimicking higher dimension. Here, more quantitatively, we may define the effective dimension $D$ as a function of the lattice dimension $d$ and the exponent $\alpha$
\begin{equation}  D = 2d / \alpha  \,,\, \end{equation} 
so that $D = d$ is recovered in the  short-range case ($\alpha=2$), and that $D = +\infty$ in the fully connected model.

\section{Positivity of soft mode}\label{sec:props}
We recall and show the following proposition that implies that the ground state of the dynamical matrix does not change sign:  if $M_{ij}$ is a square real symmetric matrix such that $M_{ij}\leq 0$ whenever $i\neq j$. Then the eigenvector $\phi = (\phi_{j})_{j=1}^L$ of its lowest eigenvalue does not change sign. 

To show this, suppose $(\phi_j)$ is normalized such that $\sum_j \phi_j^2 = 1$. Then it is known that $x_j = \phi_j$ minimizes the quadratic form $\sum_{ij} M_{ij} x_i x_j.$ In particular, 
$ \sum_{ij} M_{ij} \phi_i \phi_j \leq \sum_{ij} M_{ij} \abs{\phi_i} \abs{\phi_j} \,. $
Canceling out the diagonal terms, and using the fact that $M_{ij} < 0$ when $i \neq j$, we deduce that $\phi_i \phi_j \geq \abs{\phi_i} \abs{\phi_j} \Rightarrow \phi_i \phi_j \geq 0$ for any pair $i \neq j$.

\section{Generalized Anderson model: analogy between the linearized Riccati analysis and the Larkin approach}\label{sec:riccati}
In this section, we propose a simple interpretation of the results obtained in section \ref{sec:SR}, in the collective pinning (small disorder) regime. In particular, we show that the decay property of the ground state $\phi_j$ of the Anderson model is directly related to the interface roughness of the Larkin model, reviewed in section \ref{sec:Larkin}. Our approach is directly inspired by the Riccati-equation analysis commonly used for studying Anderson localization in 1d.

Let $\phi_j$ be the ground state of $M_{ij} = -G_{ij} + \delta_{ij} W_j$ with eigenvalue $E$. By definition, $\phi_j$ satisfies the following Equation:
 \begin{equation}
\sum_{\abs{k - j} = 1} G_{kj} (\phi_k - \phi_j) = (W_j - E) \phi_j  \,, \label{eq:eigen_gen}
 \end{equation}
Since $\phi_j$ does not change sign, we may suppose $\phi_j > 0$ for all $j$ and denote 
\begin{equation}
 v_j = \ln \phi_j  \Leftrightarrow \phi_j = e^{v_j} \,, \label{eq:colehopf}
\end{equation}
in terms of which Eq. \eqref{eq:eigen_gen} becomes 
\begin{equation}
\sum_{\abs{k - j} = 1} G_{kj} (e^{v_k - v_j} - 1) = W_j - E \,. \label{eq:eigen_gencolehopf} 
\end{equation}

To proceed further, let us assume that the disorder is weak, $\abs{W_j - E} \sim \sigma / r_f \ll G_{jk} \sim 1$ [see Eq. \eqref{eq:Wdmatch}]. Then, Eq. \eqref{eq:eigen_gen} suggests the working hypothesis that the magnitudes of $\phi_j$ are slowly varying, that is,
\begin{equation}
\abs{\phi_{j\pm1} - \phi_j} \ll \abs{\phi_j} \Leftrightarrow \abs{v_{j \pm1}  - v_j} \ll 1 \,, \label{eq:slowhypothesis}
\end{equation}
Under this hypothesis, we may expand the exponential in Eq. \eqref{eq:eigen_gencolehopf}:
\begin{equation}
\sum_{\abs{k - j} = 1} G_{kj} \left[v_k - v_j  + \frac{(v_k - v_j)^2}{2} + \dots\right] = W_j - E \,. \label{eq:Riccatidiscrete}
\end{equation}

Remark that by taking the continuum limit of Eq. \eqref{eq:Riccatidiscrete}: $v_j = v(j \delta x), W_j = U(j \delta x)$, with $\delta x \to 0$, we obtain the Riccati Equation:
\begin{equation}
\frac{\dif z(x)}{\dif x} + z(x)^2 = U(x) - E \,,\, z(x) = v'(x) \,. \label{eq:Riccati}
\end{equation}
Usually this Equation is used to calculate the (complex) Lyapunov exponent $\Omega(E)$ which describes well the localization properties in the bulk (for example, one can show that $\xi_{\text{loc}} \propto \sigma^{-2}$, see Fig.~\ref{fig:eigenvectors} above). Yet, the localization properties of the ground state is \textit{not} given by $\Omega(E)$ in a simple way~\cite{Comtet2010,texier10dirac,comtet13lyapunov}.
To proceed further, we propose to linearize the Riccati Equation. In the discrete setting, this amounts to neglecting the quadratic term in Eq. \eqref{eq:Riccatidiscrete}, which is justified by Eq. \eqref{eq:slowhypothesis} near the localization center. The resulting Equation
\begin{equation}  \sum_{\abs{k - j} = 1} G_{kj} \left(v_k - v_j\right) = W_j - E \,, \label{eq:RAP}  \end{equation}
is formally identical to the Larkin model, Eq. \eqref{eq:Larkinmodel}, upon replacing $v_j \to \tilde{u}_j$ and $W_j \to \left<  F \right> - F_j$. Therefore, the analysis in section \ref{sec:Larkin} applies (with $d = 1, \alpha = 2$). Recalling the magnitude of the random ``forces'' $W_j - E \sim \sigma / r_f$, we have 
\begin{equation}
\overline{(v_j - v_k)^2} \sim (\sigma / r_f)^2 \abs{j-k}^{3} \,. 
\end{equation}
Therefore, $v_j$ is a random acceleration process, and the ground state is its exponential $\phi_j = e^{v_j}$. In particular, since $v_j$ is a rough surface with one dominating maximum region $j \sim j_{\max}$, $\phi_j$ is localized around $j_{\max}$, and has a stretch exponential decay 
\begin{equation}
\phi_j \sim \exp\left(- \abs{\frac{j-j_{\max}}{\ell_c}}^{\frac32}  \right) \,,\, \ell_c \sim \left(\frac{r_f}{\sigma}\right)^{\frac23} \,, \label{eq:stretchexpo}
\end{equation}
where we note that the corresponding localization length $\ell_c$ is \textit{identical} to the Larkin length in Eq. \eqref{eq:Larkin}. The assumption $r_f/\sigma \ll 1$ is Equivalent to $\ell_c \gg 1$, corresponding to the collective pinning regime (see Figure \ref{fig:phase}). This result provides a clear explanation of the numerical results in section \ref{sec:SR}. Note that our linearizing approximation cannot account for the exponential far tail, which should be an effect of the non-linear term in Eq. \eqref{eq:Riccati}. 

In Ref.~\cite{Tanguy2004}, it was already observed that the localization length of the ground states satisfies the power-law $\xi_{\text{loc}}\sim \sigma^{-2/3}$, as long as $\xi_{\text{loc}} < L$. Thus, for smaller disorder, the ground state appears to be delocalized. This was interpreted as a ``weak pinning'' regime, which we prefer to call a finite-size crossover, since the delocalization disappears as $L \to \infty$.

\begin{figure}

\includegraphics[width=.97\columnwidth,valign=t]{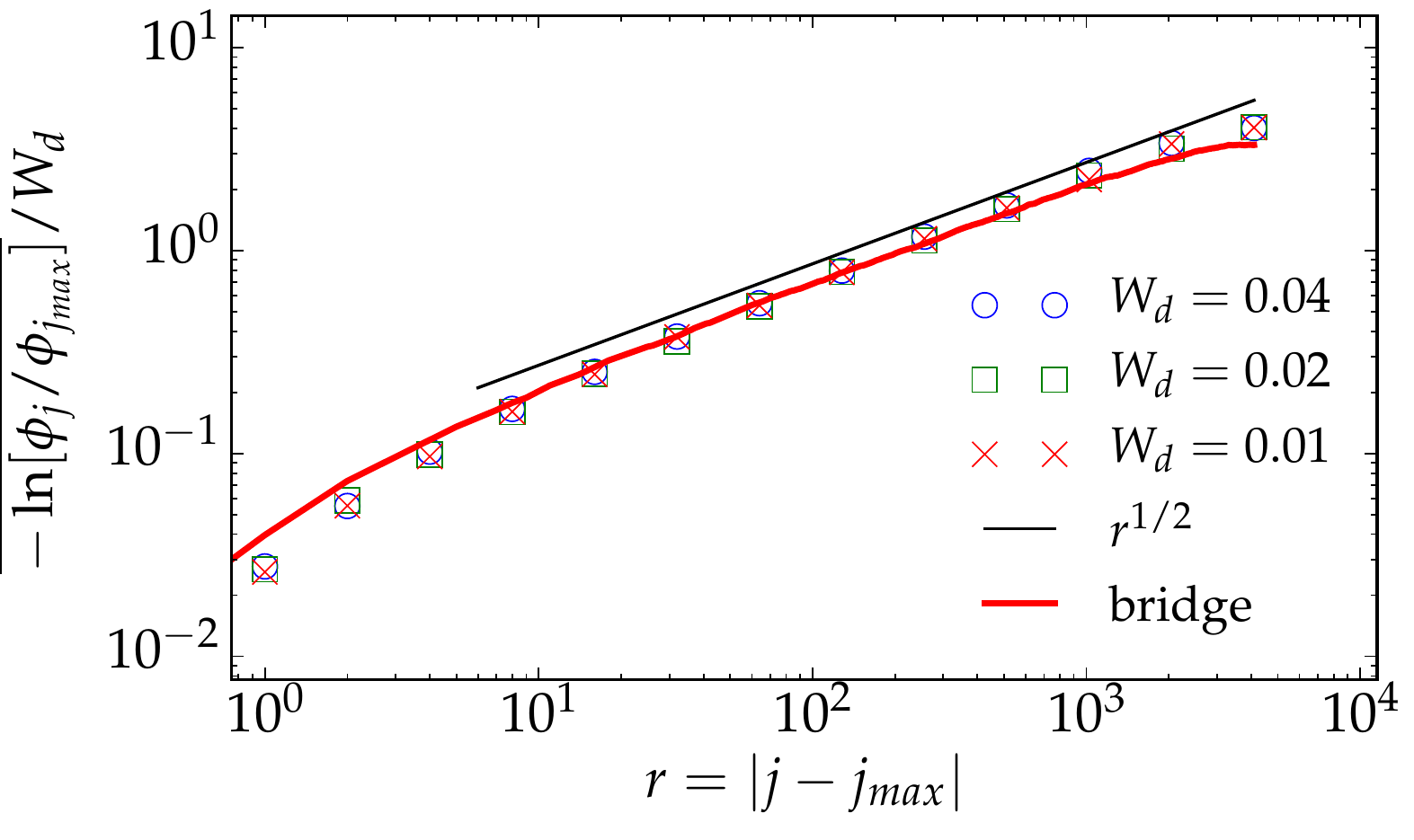}
\caption{Numerical measure of the decay of ground state of the 1d long-range Anderson model with $\alpha = 1$, compared to the prediction Eq. \eqref{eq:stretchgeneral}. The diagonal elements are uniformly distributed in $[0,W_d]$. Close to the maximum $j_{\max}$ , we observe a small deviation from the $r^{1/2}$ power-law. A similar behavior is also observed around the maximum of a Brownian bridge (red curve). }\label{fig:anderground}
\end{figure}

Following the discussion in Ref.~\cite{Tanguy2004}, we now study the long-range case  ($0< \alpha < 2$). Our results suggest that the depinning of an interface with long-range interaction could be studied by considering the following long-range generalization of Anderson model of Eq.\ref{eq:rodriguez}
with $\alpha \in (0, 2)$. Then, Eqs. \eqref{eq:colehopf} through \eqref{eq:eigen_gencolehopf} can be extended straightforwardly, giving:
$$\sum_{k \neq j} G_{kj} \, (e^{v_k - v_j} - 1) = W_j - E \,.$$
Using again the linearizing approximation $e^{v_k - v_j} - 1 \approx v_k - v_j$, we obtain the following generalization of Eq. \eqref{eq:RAP}:
\begin{equation}  \sum_{k \neq j} G_{kj} \, (v_k - v_j) = W_j - E \,. \label{eq:LarkingenRiccati} \end{equation}
This can be compared to the generalized Larkin model, Eq. \eqref{eq:Larkinmodel}. Therefore, by adapting the above reasoning, we obtain the following prediction: when $D = 2 d  / \alpha < D_{uc} = 4$, the interface $v_j$ becomes rough (\textit{i.e.}, has large height fluctuation) at large enough distance. Accordingly, $\phi_j$ is localized around a center $j_{\max}$, with a stretched-exponential decay:
\begin{equation}
\phi_j \sim \exp\left(-\abs{\frac{j-j_{\max}}{\ell_c}}^{\zeta_L} \right)  \,,\, \label{eq:stretchgeneral}
\end{equation}
where $\ell_c = \left( \frac{r_f}{\sigma}\right)^{1/\zeta_L}$ is the Larkin length defined in Eq. \eqref{eq:Larkin}, and $\zeta_L =  \alpha-d/2 > 0$ is the Larkin exponent defined in Eq. \eqref{eq:Larkinlenght}. A test of the prediction for $d = 1$ and $\alpha = 1$ (relevant for the crack front of fracture~\cite{gao1989first} and the wetting line in the liquid meniscus of a rough substrate~\cite{joanny1984model}) is given in Figure \ref{fig:anderground}. Again, we stress that Eq. \eqref{eq:stretchgeneral} is valid only near the localization center and in the collective pinning regime. Similarly to the short-range case, to describe the far tail, one should look at a single impurity problem, and we expect a power-law decay $\phi(x) \sim \abs{x}^{-\alpha-d}$.

\section{Fully connected depinning: the dynamical matrix}\label{sec:fullymath}
Here we give elementary properties of the fully connected dynamical matrix 
    \begin{equation}
    H_{ij} = - \frac{1}L + \delta_{ij} W_j \,,\, 1\leq i,j \leq L \,.
    \end{equation}
    By a permutation of indices we may assume $W_1 \leq W_2 \leq W_3 \leq \dots \leq W_L$. 
    Let $\lambda$ be an eigenvalue of $H_{ij}$ and $(\phi_j)$ its eigenvector. Then we have 
    \begin{eqnarray}
   (W_j - \lambda) \phi_j =  \langle \phi \rangle \,,\, \langle \phi \rangle = \frac{1}L \sum_{j=1}^L \phi_j \,.\label{eq:eigen} 
    \end{eqnarray}
    There are two cases. 
    \begin{itemize}
     \item (Generic case) $\lambda \neq W_j$ for any $j$. Thus we can divide \eqref{eq:eigen} by $W_j - \lambda$, and sum over $j$:
      \begin{eqnarray}
      \frac{\phi_j}{\langle \phi \rangle}  = \frac{1}{W_j - \lambda} \,,\, \frac{1}L \sum_{j=1}^L\frac{1}{W_j-\lambda} = 1 \,. \label{eq:eigen_generic}
      \end{eqnarray}
      Note that $\lambda \neq W_j$ implies $\langle \phi \rangle$ cannot be zero now lest $\phi(j)$ vanishes as a vector. 
      The eigenvalue Equation \eqref{eq:eigen_generic} has exactly one solution in the interval $(-\infty, W_1)$ (this is the ground state), as well as in each of the intervals $(W_{j}, W_{j+1})$ when $W_{j}< W_{j+1}$ (this is the $j$-th excited state). By \eqref{eq:eigen_generic}, the ground state does not changes sign, while the excited states change sign once.
    \item (Degenerate case) $\lambda = W_j$ for some $j$, then $\langle \phi \rangle = 0$. So $(W_j - \lambda) \phi(j) = 0$ for all $j$. 
       Yet $\phi(j)$ can not be identically zero, moreover $\langle \phi \rangle = 0$ entails at least two $\phi(j)$ must be non-zero, and for these indices $W_j = \lambda$. Suppose $\lambda = W_{j-1} < W_{j} \dots = \dots W_{j'} < W_{j'+1}$, $j < j'$. Then 
       the eigenvalue $\lambda$ has $j' - j$-dimensional eigen-space, spanned by vectors of type $(0, \dots, 0, 1, -1, 0, \dots, 0)^t$ where the two non-zero coefficients occur at $i$ and $i+1$, for $i = j, \dots, j'-1$.
     \end{itemize} 
      In summary, the eigenvalues and $W_j$ satisfy the interlacing relation 
      \begin{eqnarray}
      \lambda_0 < W_1 \leq \lambda_1 \leq W_2 \leq \dots \leq \lambda_{N-1} \leq W_N \,, \label{eq:interlacing}
      \end{eqnarray}
      where Equality between $\lambda_j$ and $W_{j,j+1}$ can only happen when $W_j = W_{j+1}$. 
      
    Eq. \eqref{eq:eigen_generic} gives the exact expression for the eigenstates in terms of $W_j$. So the localization properties can be studied quite straightforwardly once we know the distribution of $W_j$. The main text will focus on the ground state, which is never degenerate. In particular, marginal stability condition $\lambda_0 = 0$ is Equivalent to  
     \begin{equation}
     \sum_j \frac1{W_j} = L \,.
     \end{equation}

\subsection{Critical force}\label{sec:fc>0}
We review in more detail the calculation of the critical force $f_c$ in the fully connected model. Its general expression is given in  Ref.~\cite{fisher1983threshold,fisher85}, and writes as:
\begin{equation}
f_c = \frac{1}{r_f}\int_{-r_f/2}^{r_f/2}  \sigma h(v) \frac{\dif P( v_j < v)}{\dif v}  \dif v \,,
\end{equation} 
where $P(v_j < v)$ is the distribution given by Eq. \eqref{eq:Pu}. Thus, for $\sigma<\sigma_c$ we get a vanishing critical force 
\begin{align}
f_c = \frac{-\sigma}{r_f}  \int_{-r_f/2}^{r_f/2} (1-\sigma\, h'(v))  h(v) \dif v=0 \,.
\end{align}
When $\sigma > \sigma_c$, the critical force writes as follows
\begin{align} 
f_c &= -\frac{\sigma}{r_f} \int_{v_+}^{r_f/2 + v_-}   (1 - \sigma h'(v)) h(v) \dif v \\ 
& = \frac{\sigma}{r_f} \int_{v_-}^{v_+}  (1 - \sigma h'(v)) h(v) \dif v \,. \label{eq:fc}
\end{align}
It is an easy exercise to show that $f_c> 0$ in general for $\sigma > 1$. For the quadratic force profile Eq. \eqref{eq:fquadratic}, the above Equation can be explicitly evaluated, giving Eq. \eqref{eq:fcquadratic}. 

In the rest of the appendix, we consider the behavior of $f_c$ near the critical disorder, 
\begin{equation} \sigma = 1 + \epsilon \,,\,  0 < \epsilon \ll 1 \,. \end{equation}
It turns out that the result is sensitive to the analyticity of $f(v)$ near $v = 0$, and is different for the two force profiles that we consider. To treat both cases in one setting, we introduce an interpolating family of force profiles, characterized by their analytical behavior near $v = 0$, and parametrised by an exponent $\eta$: 
\begin{equation} f(v) = v - a v \abs{v}^\eta + o(\abs{v}^{\eta+1}) \,. \label{eq:feta} \end{equation} 
In particular, the quadratic profile Eq. \eqref{eq:fquadratic} corresponds to $\eta = 1$ and  $f(u) = \sin u$ has $\eta = 2$. Using Eq. \eqref{eq:shock} and \eqref{eq:fc}, it is not hard to obtain the following:
\begin{equation}
v_\pm \propto \epsilon^{\frac1\eta} \,,\, \alpha_c \propto  \epsilon^{1 + \frac1\eta} \,,\, 
f_c \propto v_\pm \alpha_c = \epsilon^{1 + \frac{2}{\eta}} \,. \label{eq:critforcesummary}
\end{equation}

\bibliography{DES}

\end{document}